\documentclass[reprint,floatfix]{revtex4-2}
\usepackage{graphicx}
\usepackage{bm}
\usepackage{ulem}
\usepackage{xcolor}
\usepackage{amsmath}
\usepackage{amssymb}
\usepackage{float}
\usepackage{physics}
\usepackage{siunitx}
\usepackage{bm}
\usepackage[space]{grffile}

\makeatletter
\renewcommand{\fnum@figure}{\textbf{Figure \thefigure}}
\makeatother

\begin{document}

\title{Autonomous quantum error correction of Gottesman-Kitaev-Preskill states}
\date{\today}
\author{Dany Lachance-Quirion}
\email{dany@nordquantique.ca}
\author{Marc-Antoine Lemonde}
\author{Jean Olivier Simoneau}
\author{Lucas St-Jean}
\author{Pascal Lemieux}
\author{Sara Turcotte}
\author{Wyatt Wright}
\author{Amélie Lacroix}
\author{Jo\"elle Fréchette-Viens}
\author{Ross Shillito}
\author{Florian Hopfmueller}
\author{Maxime Tremblay}
\author{Nicholas E. Frattini}
\author{Julien Camirand Lemyre}
\author{Philippe St-Jean}
\affiliation{Nord Quantique, Sherbrooke, Québec, J1J 2E2, Canada}

\begin{abstract}
The Gottesman-Kitaev-Preskill (GKP) code encodes a logical qubit into a bosonic system with resilience against single-photon loss, the predominant error in most bosonic systems. Here we present experimental results demonstrating quantum error correction of GKP states based on reservoir engineering of a superconducting device. Error correction is made autonomous through an unconditional reset of an auxiliary transmon qubit. The lifetime of the logical qubit is shown to be increased from quantum error correction, therefore reaching the point at which more errors are corrected than generated.
\end{abstract}

\maketitle
Improving error correction schemes is a central challenge towards the development of fault-tolerant quantum processors. A high-quality bosonic mode controlled by an auxiliary nonlinear element has proven a valid candidate to replace the standard two-level system approach\,\cite{Krinner2022,Zhao2022,Sundaresan2023,GoogleQuantumAI2023}, bringing to life the visionary proposal of Gottesman, Kitaev and Preskill for error correction at the individual qubit level\,\cite{Gottesman2001,Albert2018,Noh2020,Ma2021,Cai2021,Grimsmo2021,Brady2023}. This approach is gaining momentum, thanks in part to recent experiments showing an increase in logical lifetime of bosonic codes through error correction\,\cite{Ofek2016,Hu2019,Campagne-Ibarcq2019a,Gertler2021,deNeeve2022,Sivak2023,Ni2023}, and promises to ease the requirements on number of modes needed for useful quantum computation\,\cite{Noh2020,Ma2021,Cai2021}.

Most of the recent bosonic code error correction experiments in superconducting circuits have relied on measurements of the auxiliary nonlinear element to condition real-time feedforward operations\,\cite{Ofek2016,Hu2019,Campagne-Ibarcq2019a,Sivak2023,Ni2023}, thereby introducing challenges related to measurement fidelity and complexity and latency of control electronics. Fully autonomous quantum error correction protocols for bosonic codes alleviate those challenges by reducing reliance on measurement of the auxiliary\,\cite{Barnes2000,Lihm2018,Gertler2021,deNeeve2022,Zeng2023}.

Recent work introduced more efficient protocols for error correction of finite-energy Gottesman-Kitaev-Preskill (GKP) qubits\,\cite{Royer2020,deNeeve2022}. With this approach, an improvement of the logical lifetime of the GKP qubit by a factor of $2.27$ over the logical lifetime of the Fock encoding was demonstrated\,\cite{Sivak2023}. There, the reset of the auxiliary qubit, necessary for the reservoir-engineered QEC protocol, was implemented with a measurement of the auxiliary followed by feedforward operations on both the auxiliary and the bosonic mode.

In this letter, we present experimental results demonstrating a fully-autonomous QEC protocol of a GKP qubit through a feedforward-free reset of the auxiliary transmon qubit. Despite the auxiliary having 10 times stronger relaxation rate than the bosonic mode, logical lifetime of the GKP qubit is shown to increase when applying error correction. Experimental results are supported with numerical simulations which show quantitative agreement for both the initialization and the quantum error correction of GKP states.

\begin{figure*}[t]
    \centering
    \includegraphics[scale=0.81]{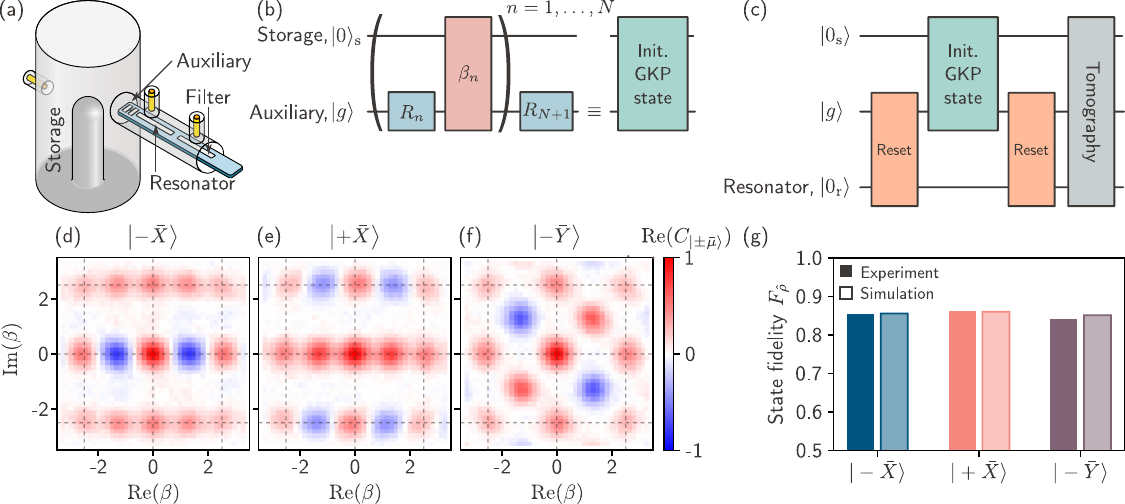}
    \caption{
    (a)~Schematic of hardware architecture. The storage mode, in which the GKP code is encoded, is the fundamental mode of a coaxial cavity dispersively coupled to an auxiliary transmon qubit. The auxiliary is also dispersively coupled to an on-chip resonator used for readout and reset.    
    (b)~Protocol for the initialization of GKP logical states with a depth-$N$ circuit. Each layer $n$ consists of an auxiliary rotation of parameter $R_n$ and a conditional displacement on the storage mode of parameter $\beta_n$.
    (c)~Protocol for measurement of the characteristic function of initialized GKP logical states.
    Real part of the characteristic function, $\mathrm{Re}(C_{\ket{\pm\bar\mu}})$, of GKP logical states (d)~$\ket{-\bar X}$, (e)~$\ket{+\bar X}$, and (f)~$\ket{-\bar Y}$ for $\Delta=0.36$ using a circuit with a depth $N=9$, leading to an initialization duration of $7.86\,\unit{\micro\second}$ excluding resets.
    (g)~State fidelity estimated from state reconstruction given measured (d-f) and simulated characteristic functions.
    }
    \label{fig:hardware_and_initialization}
\end{figure*}

A multi-element superconducting device is used to experimentally realize autonomous QEC of GKP states in a circuit quantum electrodynamics architecture\,\cite{Blais2021}. The bosonic mode is hosted inside a three-dimensional seamless coaxial cavity\,\cite{Reagor2016} made out of high-purity aluminum [Fig.\,\ref{fig:hardware_and_initialization}(a)]. The fundamental $\lambda/4$ mode of the cavity, hereafter the \textit{storage mode}, has a lifetime of $T_{1\mathrm{s}}=0.34\,\unit{\milli\second}$. A chip in the coaxline architecture is used to provide the nonlinearity required to control the storage mode\,\cite{Axline2016}. A transmon qubit, a resonator and a Purcell filter are fabricated on this silicon chip\,\cite{Koch2007,Blais2021}. The auxiliary transmon qubit has a relaxation time $T_{1\mathrm{q}}=33\,\unit{\micro\second}$ and echo coherence time $T_{2\mathrm{q}}=48\,\unit{\micro\second}$. The resonator, used for readout and reset of the auxiliary transmon\,\cite{Magnard2018,Egger2018,Sunada2022}, has a fundamental $\lambda/2$ mode that is overcoupled through the Purcell filter to an output port with total decay rate $\kappa_\mathrm{r}/2\pi=1.7$\,MHz. The chip is slightly inserted inside the microwave cavity to overlap with the storage mode, leading to a dispersive shift of $2\chi_\mathrm{sq}/2\pi=-22$\,kHz per excitation\,\cite{Reagor2016,Axline2016}.

The GKP states initialized and error corrected in this manuscript are the six finite-energy cardinal states
$\ket{\pm\bar\mu}$ with $\bar\mu\in\left\{\bar X,\bar Y,\bar Z\right\}$ of finite-energy parameter $\Delta$. As shown in Fig.\,\ref{fig:hardware_and_initialization}(b), we experimentally prepare these states by applying a circuit with alternating auxiliary rotations and conditional displacements of the storage mode, exploiting that such circuits enable approximate universal control of the storage mode\,\cite{Eickbusch2022}. Conditional displacements, which displace the state of the storage mode with a sign conditioned on the state of the auxiliary, are implemented with the echoed conditional displacement protocol in $0.85\,\unit{\micro\second}$\,\cite{Campagne-Ibarcq2019a,Eickbusch2022}. The initialization protocol parameters for the auxiliary rotation and conditional displacements of the storage mode are found using optimization for a given target state $\ket{\pm\bar\mu_0}$ and parameter $\Delta$\,\cite{SM}.

We perform tomography of the prepared states by measuring the characteristic function $C_{\ket{\pm\bar\mu}}(\beta)$ of the storage mode using the auxiliary initialized in $\ket{g}$, where $\beta$ is the amplitude of the tomography conditional displacement\,\cite{Fluhmann2020,Campagne-Ibarcq2019a,Eickbusch2022}. The complete protocol for initialization and tomography of GKP logical states is shown in Fig.\,\ref{fig:hardware_and_initialization}(c). The initial reset, whose details are discussed later, decreases to probability of the auxiliary to be in its excited state $\ket{e}$ from the thermal population of $0.4\%$ to $0.1\%$. The reset after the initialization protocol approximately ensures that the tomography protocol starts with the auxiliary in $\ket{g}$ even if errors have occurred during the initialization protocol.

The real part of the characteristic function for the GKP states $\ket{-\bar X}$, $\ket{+\bar X}$, and $\ket{-\bar Y}$ are shown in Fig.\,\ref{fig:hardware_and_initialization}(d--f) for finite-energy parameter $\Delta=0.36$, corresponding to an average number of photons $\overline{n}_\mathrm{s}=3.5$ and squeezing of $8.9$\,dB. The remaining cardinal states of the GKP qubit are obtained by a rotation in phase space with $\beta_n\rightarrow i\beta_n$. The fidelity of initialized GKP logical states is evaluated with $F_{\hat\rho}=\mathrm{Tr}\sqrt{\hat\rho^{1/2}\hat\rho_\mathrm{target}\hat\rho^{1/2}}$, where $\hat\rho$ is the density matrix obtained via state reconstruction from measurements of the characteristic function\,\cite{Strandberg2022}, and $\hat\rho_\mathrm{target}$ is the density matrix of ideal finite-energy GKP logical states [Fig.\,\ref{fig:hardware_and_initialization}(g)]. State fidelity averaged over the cardinal states reaches $85\%$. By fitting fidelities from numerical simulations to the experimentally estimated fidelities, an intrinsic dephasing of the storage mode at a rate $\kappa_\mathrm{s\phi}=(110\,\mathrm{ms})^{-1}$ is estimated, which is consistent with the value of Ref.\,\cite{Eickbusch2022}. Storage mode dephasing, amplified during large displacements\,\cite{Eickbusch2022}, explains a preparation infidelity of $6.4\%$, while decay of the auxiliary leads to an additional $4.0\%$\,\cite{SM}.

While the full tomography in phase space is useful to estimate the density matrix and state fidelity from reconstruction, the logical information of the GKP qubit can be measured more efficiently. The logical fidelity $F_\mathrm{L}$ may be expressed as
\begin{align}
F_\mathrm{L}=\frac{1}{2}+\frac{1}{12}\sum_{\hat\mu_0\in\left\{\hat X_0,\hat Y_0,\hat Z_0\right\}}\left(\expval{\hat\mu_0}_+-\expval{\hat\mu_0}_-\right),
\label{eq:logical_channel_fidelity}
\end{align}
where $\expval{\hat\mu_0}_\pm=\expval{\pm\bar\mu|\hat\mu_0|\pm\bar\mu}$ is the expectation value of the infinite-energy Pauli operator $\hat\mu_0$ when a logical state $\ket{\pm\bar\mu}$ with $\bar\mu\in\left\{\bar X,\bar Y,\bar Z\right\}$ is prepared\,\cite{Nielsen2002,Sivak2023}. The Pauli expectation value corresponds to the real part of the characteristic function $C_{\ket{\pm\bar\mu}}(\beta_{\hat\mu_0})$ of the state $\ket{\pm\bar\mu}$ with $\beta_{\hat X_0}=\ell/\sqrt{8}$, $\beta_{\hat Y_0}=(1+i)\ell/\sqrt{8}$ and $\beta_{\hat Z_0}=i\ell/\sqrt{8}$ with $\ell=2\sqrt{\pi}$ for the square GKP qubit\,\cite{Terhal2020a,Royer2020}. After state preparation, the logical fidelity reaches $84\%$.

\begin{figure}[t]
    \centering
    \includegraphics[scale=0.81]{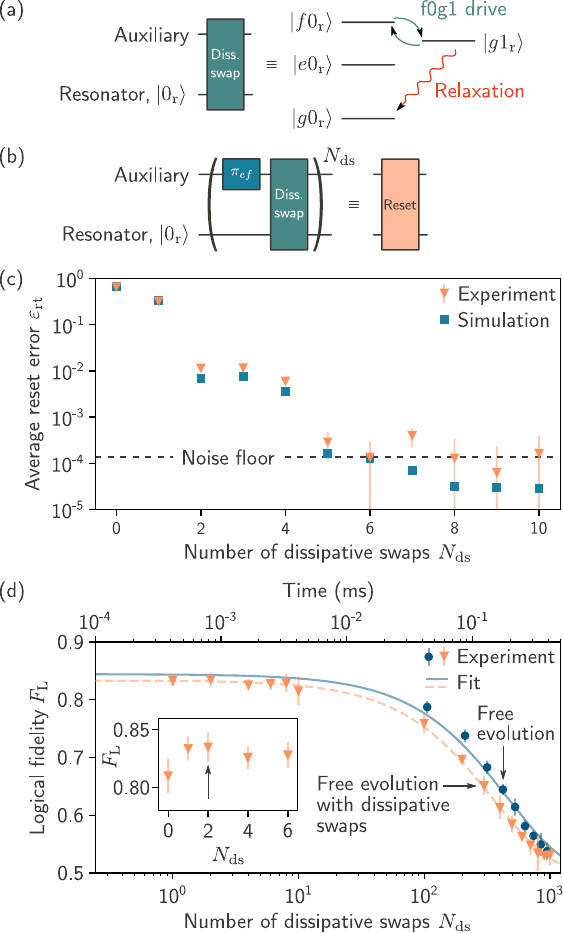}
    \caption{
    (a)~Dissipative swap between the auxiliary qubit and the resonator enabled by an effective $\ket{f0}_\mathrm{r}\leftrightarrow\ket{g1}_\mathrm{r}$ transition, followed by relaxation from $\ket{g1}_\mathrm{r}\rightarrow\ket{g0}_\mathrm{r}$.
    (b)~Reset protocol with $N_\mathrm{ds}$ dissipative swaps.
    (c)~Average reset error $\varepsilon_\mathrm{rt}$ as a function of $N_\mathrm{ds}$ measured experimentally (orange) and obtained from numerical simulations (teal). The horizontal dashed line indicates the experimental noise floor of $4\times10^{-4}$.
    (d)~Logical fidelity $F_\mathrm{L}$ measured as a function of free evolution time (blue, top axis) or number of dissipative swaps (orange, bottom axis). The plain and dashed lines show exponential fits to the data for free evolution without and with dissipative swaps, respectively. Inset: zoom to highlight the optimal value of $N_\mathrm{ds}=2$.
    }
    \label{fig:reset}
\end{figure}

A reset of the auxiliary qubit is required for error correction of GKP states based on reservoir engineering, as it constitutes the dissipative channel through which entropy is evacuated\,\cite{Royer2020}. The error correction can be made completely autonomous using a microwave-activated unconditional reset\,\cite{Magnard2018,Egger2018,Sunada2022}. As shown in Fig.\,\ref{fig:reset}(a), the reset is based on swapping two excitations from the auxiliary to a single readout resonator excitation through an effective $\ket{f0_\mathrm{r}}\leftrightarrow\ket{g1_\mathrm{r}}$ transition of the coupled system\,\cite{Zeytinoglu2015,Magnard2018,Egger2018}, where $\ket{kn_\mathrm{r}}=\ket{k}\otimes\ket{n_\mathrm{r}}$ corresponds to the qubit state $\ket{k\in\left\{g,e,f,\dots\right\}}$ and resonator Fock state $\ket{n_\mathrm{r}\in\left\{0_\mathrm{r},1_\mathrm{r},2_\mathrm{r},\dots\right\}}$. The resulting photon in the resonator dissipates to the environment through $\ket{g1_\mathrm{r}}\rightarrow\ket{g0_\mathrm{r}}$ on a timescale corresponding to resonator lifetime $T_\mathrm{1r}=1/\kappa_\mathrm{r}=92$\,ns. The complete process is an effective auxiliary-resonator dissipative swap.

To reset the first excited state $\ket{e}$ of the auxiliary, the dissipative swap is prepended with a $\pi$-pulse addressing the $\ket{e}\leftrightarrow\ket{f}$ transition [Fig.\,\ref{fig:reset}(b)]. This base sequence can be repeated to reset the second excited state $\ket{f}$ of the qubit. The reset protocol --- which takes $414$\,ns per swap including resonator decay --- is repeated $N_\mathrm{ds}$ times to decrease the reset error. Ideally, the qubit ends up in the ground state after the reset regardless of the initial state $\ket{k\in\left\{g,e,f\right\}}$ for $N_\mathrm{ds}\geq2$.

The average reset error, defined as
\begin{align}
\varepsilon_\mathrm{rt}=\frac{1}{3}\sum_{k\in\left\{g,e,f\right\}}\left(1-p_{g|k}\right),
\end{align}
where $p_{g|k}$ is the probability of the qubit being in the ground state $\ket{g}$ when preparing state $\ket{k}$, is measured as a function of the number of dissipative swaps $N_\mathrm{ds}$ with the storage mode in equilibrium [Fig.\,\ref{fig:reset}(c)]. When the storage mode is not in a Fock state $\ket{n_\mathrm{s}}$, undesired entanglement is formed between auxiliary and storage mode during the reset process due to their always-on dispersive interaction. For instance, this leads to an increase of the reset error for $N_\mathrm{ds}=2$ from $1.4\%$ in the vacuum to $2.8\%$ in the presence of a GKP logical state $\ket{-\bar X}$\,\cite{SM}.

The optimal number of dissipative swaps per reset for QEC needs to be chosen to achieve a low reset error, while avoiding any adverse effects on the GKP state. Figure\,\ref{fig:reset}(d) shows that a decrease of the logical lifetime $T_\mathrm{L}$ from $0.182$\,ms to $0.143$\,ms is observed when replacing free evolution of GKP states with a reset composed of $N_\mathrm{ds}$ dissipative swaps\,\cite{SM}. Given that the reset process should generate an identity operation on the storage mode when the auxiliary is in $\ket{g}$, the observed decrease of the logical lifetime hints at an additional dephasing mechanism whose investigation is left to future work. The inset of Fig.\,\ref{fig:reset}(d) indicates that $N_\mathrm{ds}=2$ maximizes the logical fidelity of the GKP qubit. This value is thus used throughout the manuscript.

\begin{figure}[t]
    \centering
    \includegraphics[scale=0.81]{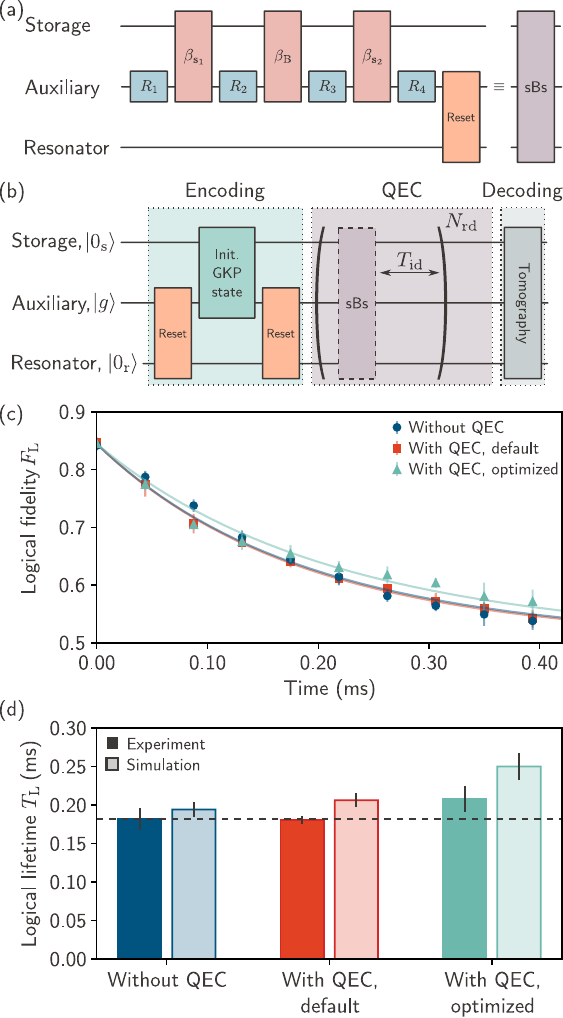}
    \caption{
    (a)~Quantum error correction protocol based on the s\textbf{B}s protocol composed of auxiliary rotations, conditional displacements and a reset of the auxiliary.
    (b)~Protocol used to measure the logical lifetime. After initialization of a GKP logical state (encoding), the s\textbf{B}s protocol is either applied (with QEC) or not (without QEC) $N_\mathrm{id}$ times. An extra idle time $T_\mathrm{id}$ between each round is added in both cases. The GKP qubit Pauli expectation value $\expval{\hat\mu_0}_\pm$ is measured for states $\ket{\pm\bar\mu}$ (decoding).
    (c)~Logical fidelity $F_\mathrm{L}$ measured as a function of time without QEC (blue), and with default (red) or optimized s\textbf{B}s protocols (turquoise) for $T_\mathrm{id}=40~\unit{\micro\second}$. The full lines show the decay of the logical fidelity obtained from a fitting procedure\,\cite{SM}. The data without QEC is the same as in Fig.\,\ref{fig:reset}(d).
    (d)~Logical lifetime $T_\mathrm{L}$ obtained experimentally (dark bars) and numerically (pale bars). The horizontal dashed line corresponds to the logical lifetime without QEC.
    }
    \label{fig:logical_lifetime}
\end{figure}

The quantum error correction protocols for GKP states proposed in Ref.\,\cite{Royer2020} engineer a set of two dissipators whose ground state manifold corresponds to the desired logical manifold even in the presence of single-photon loss in the storage mode. A single round of the small-Big-small (s\textbf{B}s) protocol approximates each of the required dissipators by a depth-$3$ circuit of auxiliary rotations and conditional displacements, followed by a dissipative operation implemented through unconditional reset of the auxiliary [Fig.\,\ref{fig:logical_lifetime}(a)]. The parameters of the s\textbf{B}s protocol are $\bm{\beta}_\mathrm{s\textbf{B}s}=\left(\beta_{\mathrm{s}_1},\beta_\mathbf{B},\beta_{\mathrm{s}_2}\right)$ with
\begin{align}
\beta_{\mathrm{s}_{1,2}}^{(n)}&=\frac{i^{(n-1)}\ell\sinh\Delta_\mathrm{s\textbf{B}s}^2}{2\sqrt{2}},\hspace{0.2cm}\beta_\mathbf{B}^{(n)}=\frac{i^n\ell\cosh\Delta_\mathrm{s\textbf{B}s}^2}{\sqrt{2}},
\label{eq:stabilization_parameters}
\end{align}
for round $n$ and $\mathbf{R}_\mathrm{s\textbf{B}s}=\left(+\frac{i\pi}{2},+\frac{\pi}{2},-\frac{\pi}{2},-\frac{i\pi}{2}\right)$\,\cite{Royer2020}. The effective finite-energy parameter $\Delta_\mathrm{s\textbf{B}s}$ is used as an optimization parameter with a default value of $\Delta$. After each round, the conditional displacements are rotated a quarter turn via the $i^n$ factor, in order to implement both required dissipators while symmetrizing the conditional displacements signs.

Our main experiment measures the logical lifetime of the quantum memory. The protocol, shown in Fig.\,\ref{fig:logical_lifetime}(b), is composed of an encoding step through the initialization protocol [Fig.\,\ref{fig:hardware_and_initialization}(c)], followed by $N_\mathrm{rd}$ rounds of quantum error correction with the s\textbf{B}s protocol [Fig.\,\ref{fig:logical_lifetime}(a)], and finally a decoding step through measurement of the Pauli expectation value of the GKP qubit. An idle time $T_\mathrm{id}$ is inserted after each round of error correction in order to balance correctable single-photon loss from the storage mode and uncorrectable errors introduced by auxiliary decay during the s\textbf{B}s protocol\,\cite{Ofek2016}. The duration of the QEC step is given by $T=N_\mathrm{rd}T_\mathrm{rd}$, with time per round $T_\mathrm{rd}=T_\mathrm{s\textbf{B}s}+T_\mathrm{id}$, with $T_\mathrm{s\textbf{B}s}=3.738\,\unit{\micro\second}$ and $T_\mathrm{id}=40\,\unit{\micro\second}$. This corresponds to a photon loss probability per round $\kappa_\mathrm{s}T_\mathrm{s\textbf{B}s}=1.1\times10^{-2}$ and $\kappa_\mathrm{s}T_\mathrm{rd}=1.3\times10^{-1}$.

Figure~\ref{fig:logical_lifetime}(c) shows the logical fidelity measured experimentally without and with QEC, both with default and optimized s\textbf{B}s protocols. The optimized protocol is obtained by maximizing logical fidelity at $N_\mathrm{rd}=4$ in a closed-loop optimization. The optimization parameters are the effective finite-energy parameter $\Delta_\mathrm{s\textbf{B}s}$, the scaling of the second small displacement, parameterized by the ratio $\left|\beta_{\mathrm{s}_2}\right|/\left|\beta_{\mathrm{s}_1}\right|$, and a state rotation per round\,\cite{SM}. The optimized ratio $\left|\beta_{\mathrm{s}_2}\right|/\left|\beta_{\mathrm{s}_1}\right|=1.82$ is found to be significantly different from the value of the default protocol, yet consistent with results in Ref.\,\cite{Sivak2023}.

Figure~\ref{fig:logical_lifetime}(d) shows the logical lifetime obtained experimentally and from numerical simulations from the decay of the logical fidelity\,\cite{SM}. The gap between experimental and simulation values, when not allowing any fitting parameters, indicates that some effects are not captured in simulation. The logical lifetime is increased by $15\%$ through closed-loop optimization. Most importantly, the logical lifetime is increased by $14\%$ when applying the optimized QEC protocol compared to the lifetime of the same states without QEC, thus demonstrating that our autonomous error correction protocol corrects more errors than it generates. The logical lifetime with optimized QEC, $T_\mathrm{L}^\mathrm{opt}=0.208(17)\,\unit{\milli\second}$, is nevertheless lower than for the Fock encoding $\left\{0,1\right\}$ with $T_\mathrm{L}^\mathrm{Fock}=0.482\,\unit{\milli\second}$ in simulations\,\cite{SM}.

The gain from QEC in our experiment is limited by bit flips of the auxiliary transmon during conditional displacements of the s\textbf{B}s protocol, which can cause logical errors not protected against by the GKP code\,\cite{Campagne-Ibarcq2019a,Royer2020,Sivak2023}. The presence of this error channel forces the insertion of idle times $T_\mathrm{id}\gg T_\mathrm{s\mathbf{B}s}$ for optimal performance, which in turn limits the rate of error correction and the protection against single-photon loss in the storage mode. Both aspects call for reducing the rate of auxiliary errors with longer-lived transmons\,\cite{Place2021,Sivak2023} or fault-tolerant error syndrome extraction\,\cite{Puri2019a} with noise-biased auxiliaries\,\cite{Mirrahimi2014, Lescanne2019,Grimm2020,Frattini2022,Reglade2023}, and reducing their impact on logical states through more robust encoding of GKP qubits\,\cite{Royer2022}.

Irrespective of the error rate of the auxiliary, reducing the probability of single-photon loss during the s\textbf{B}s protocol, $\kappa_\mathrm{s}T_\mathrm{s\textbf{B}s}=1.1\times10^{-2}$ here, would increase the gain from QEC\,\cite{Royer2020}. For a fixed single-photon loss rate $\kappa_\mathrm{s}$, improving $\kappa_\mathrm{s}T_\mathrm{s\textbf{B}s}$ would require faster conditional displacements, either through optimization of the echoed conditional displacement protocol\,\cite{Eickbusch2022} or through alternative methods to implement conditional displacements in superconducting circuits\,\cite{Didier2015,Eddins2018,Touzard2019,Ikonen2019,Grimm2020,Frattini2022,Diringer2023}. Finally, the advantages of the fully autonomous approach presented here and the measure-and-feedforward approach of Ref.\,\cite{Sivak2023} could be combined. Indeed, by using a measurement followed by an unconditional reset of the auxiliary, one would have access to error syndromes that can be used in a second layer of quantum error correction, without requiring fast feedforward operations within the first layer.

In conclusion, GKP logical states have been initialized in the mode of a superconducting cavity with state fidelities mainly limited by bit flips of the auxiliary and intrinsic dephasing of the storage mode. A quantum error correcting protocol based on reservoir engineering, in which the reset of the auxiliary serves as the dissipative process, is demonstrated. The QEC protocol is made autonomous through implementation of an unconditional auxiliary reset based on a dissipative swap to a lossy resonator. Despite the auxiliary having a relaxation time $10$ times shorter than the storage mode, autonomous QEC of GKP states with a gain on the logical lifetime of about $14\%$ is demonstrated. To guide the requirements at the second layer of error correction, further work is needed to upper bound the gain from QEC of the GKP code accessible with realistic hardware and software improvements.

The authors would like to thank Alexandre Blais, Michel H. Devoret, Baptiste Royer, Alec Eickbusch, Volodymyr V. Sivak, Mathieu Juan, and Yasunobu Nakamura for useful discussions and insights. The authors would also like to thank the RIKEN Center for Quantum Computing for the fabrication of the auxiliary superconducting devices, Nature Alu for providing high-purity aluminum for the coaxial cavities, and Institut quantique of Université de Sherbrooke for access to their infrastructure.

\end{document}


\title{Supplementary Materials for "Autonomous quantum error correction of Gottesman-Kitaev-Preskill states"}
\date{\today}
\author{Dany Lachance-Quirion}
\email{dany@nordquantique.ca}
\author{Marc-Antoine Lemonde}
\author{Jean Olivier Simoneau}
\author{Lucas St-Jean}
\author{Pascal Lemieux}
\author{Sara Turcotte}
\author{Wyatt Wright}
\author{Amélie Lacroix}
\author{Jo\"elle Fréchette-Viens}
\author{Ross Shillito}
\author{Florian Hopfmueller}
\author{Maxime Tremblay}
\author{Nicholas E. Frattini}
\author{Julien Camirand Lemyre}
\author{Philippe St-Jean}
\affiliation{Nord Quantique, Sherbrooke, Québec, J1J 2E2, Canada}

\maketitle
\tableofcontents
\newpage

\section{Quantum hardware}

    \subsection{Experimental setup}

\begin{figure*}[t]
            \centering
            \includegraphics[scale=0.9]{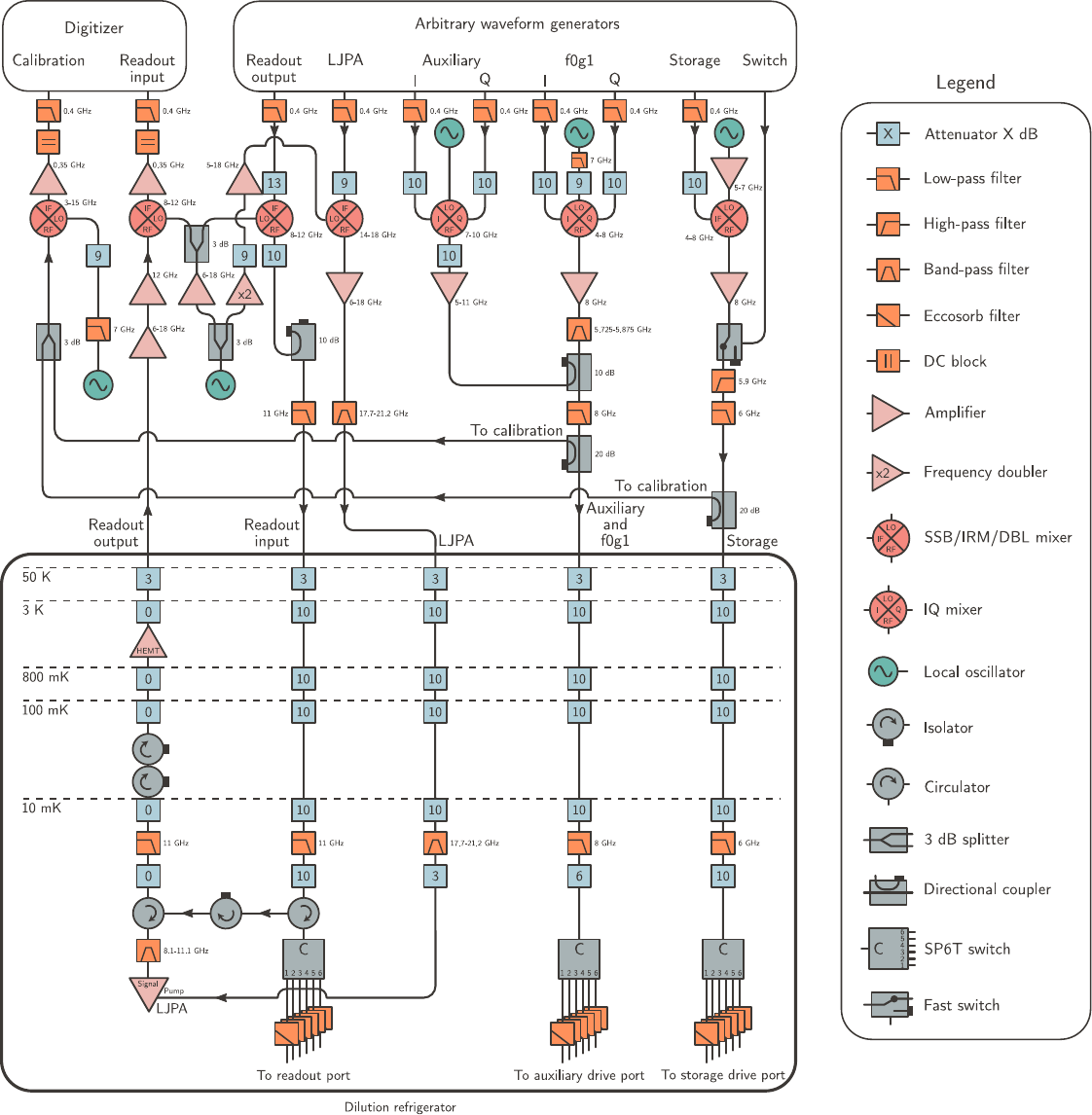}
            \caption{
            \textbf{Experimental setup used for the autonomous QEC of GKP states.}
            Signals from local oscillators and arbitrary-waveform generators are mixed with single-sideband (SSB) and in-phase quadrature (IQ) mixers. The output signal is down-converted using an image-reject mixer (IRM) with the same local oscillator as used to generate the readout pulse. A calibration setup is used to seamlessly perform or check mixer calibration without any changes to the setup. The LJPA and the corresponding pump line present in the setup are not used in the present experiments.
            }
            \label{fig:experimental_setup}
        \end{figure*}

        The setup used for the experiments is shown in Fig.\,\ref{fig:experimental_setup}. Readout and control signals are generated using up-conversion of the pulse sequences generated at $1$\,GS/s with arbitrary waveform generators (\textit{Keysight M3202A}) to the desired frequencies by mixing with local oscillators using single-sideband (SSB) and in-phase quadrature (IQ) mixers. The up-converted storage mode control signal is gated with a fast microwave switch to further suppress leakage. Signals of the auxiliary and f0g1 control channels are combined at room temperature with a directional coupler. Directional couplers are also used to tap the control signals to seamlessly check or perform mixer calibration with a dedicated calibration setup.

        Microwave switches are used to select one out of six cavity assemblies. These cavity assemblies are placed in a three-layer magnetic shield (from outside to inside: cryoperm, niobium, and cryoperm). Eccosorb filters are used on all control lines inside the magnetic shield, as close as possible to the cavity assemblies. The signal reflected on the readout port is amplified with a high-electron mobility transistor (HEMT) amplifier at $3$\,K and additional amplifiers at room temperature. A flux-pump lumped-element Josephson parametric amplifier (LJPA, \textit{RIKEN RQC}) is present on the outline line, but not used. It is worth noting that, as discussed in Sec.\,\ref{ssec:readout}, despite the absence of a parametric amplifier, a readout fidelity above $99\%$ is achieved. Down-conversion is performed using the same local oscillator as the up-conversion of the readout signal with a image-reject mixer (IRM). Further amplification of the intermediate-frequency signal is used before being digitized with a $2$\,ns resolution (\textit{Keysight M3102A}). All instruments are synchronized with a $10$\,MHz reference from a rubidium clock.

    \subsection{Information about the hardware architecture}

\begin{figure*}[t]
            \centering
            \includegraphics[scale=0.81]{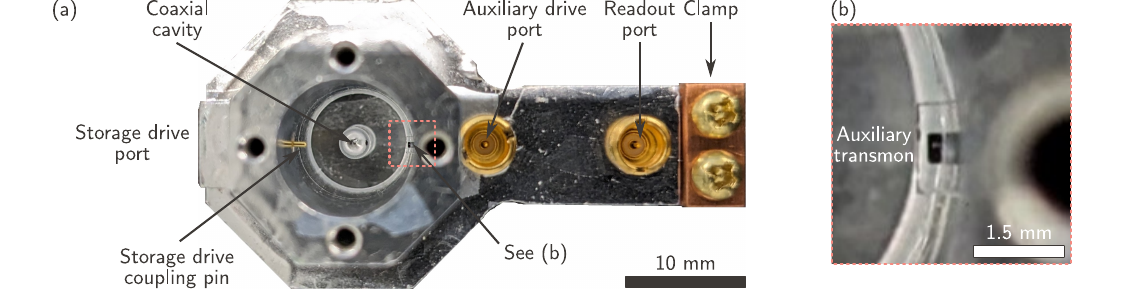}
            \caption{
            \textbf{Hardware architecture.}
            (a)~Photo of a hardware assembly nominally identical to the one used for the experiments presented in this manuscript. A coaxial cavity is machined inside a block of a high-purity aluminum. The chip, containing the auxiliary transmon, the readout and reset resonator and the Purcell filter is held in a waveguide with a copper clamp. The system is driven through the readout port, the auxiliary drive port, and the storage drive port. The insertion of the later is exaggerated to make the coupling pin visible.
            (b)~Zoom-in from (a) (pink rectangle) in which part of the auxiliary transmon is visible. Here the chip is in sapphire as opposed to the silicon chip used in the present experiments.
            }
            \label{fig:device_architecture}
        \end{figure*}

        A photo of a hardware assembly nominally identical to the one used in the experiments is shown in Fig.\,\ref{fig:device_architecture}. The cavity is machined in a block of high purity aluminum (4N+) from \textit{Nature Alu}. A standard etch process, described in Ref.\,\cite{Reagor2015}, is used to remove from all surfaces a $\sim120\,\unit{\micro\meter}$ layer damaged during machining. The relaxation time of the storage mode reached about $1.1$\,ms right after etch, but degraded to $0.34$\,ms (Sec.\,\ref{ssec:storage_relaxation}) over the two years between the initial etch and the current experiments.

        The auxiliary transmon, readout and reset resonator and Purcell filter are fabricated on a silicon chip inserted in a waveguide leading to the storage mode (Fig.\,\ref{fig:device_architecture}). All the relevant modes have frequencies between $6$ and $9$\,GHz. The chip is clamped between two pieces of copper thermalized to a copper plate with a copper thread. The cavity assemblies are thermalized to the same copper plate. An indium seal is used on the aluminum cover and aluminum Mylar tape is used to cover all seams to reduce radiation getting to the auxiliary transmon, which was observed to be critical in reducing the equilibrium population from larger than $10\%$ to less than $1\%$.

        Sub miniature push-on (SMP) connectors are used on the three ports of the assembly: the readout port (readout signal), auxiliary drive port (auxiliary and f0g1 control signals), and storage drive port (storage control). External coupling rates of these ports to the relevant modes are adjusted with precision spacers between the connectors and the cavity assembly.

\section{Theoretical framework}
    
    \subsection{Coherent dynamics and Hamiltonian parameters}

        The starting point adopted to describe the coherent dynamics taking place in the experiments is the Jaynes-Cumming (JC) Hamiltonian. Focusing on the three modes of interest, it reads 
        \begin{align}
            \hat H_\mathrm{JC}/\hbar&=\underbrace{\vphantom{\frac{K_{\mathrm{q}0}}{2}}\omega_\mathrm{s0}\hat a^\dagger\hat a+\omega_\mathrm{q0}\hat b^\dagger\hat b+\omega_\mathrm{r0}\hat c^\dagger\hat c}_\text{bare modes}+\underbrace{\frac{K_{\mathrm{q}0}}{2}\hat b^{\dagger2}\hat b^2}_\text{self-Kerr}+\underbrace{\vphantom{\frac{K_{\mathrm{q}0}}{2}}g_\mathrm{sq}\left(\hat a^\dagger\hat b+\hat a\hat b^\dagger\right)+g_\mathrm{qr}\left(\hat c^\dagger\hat b+\hat c\hat b^\dagger\right)}_\text{electric-dipole coupling},
            \label{eq:bare_Hamiltonian}
        \end{align}
        where $\hat a$, $\hat b$, and $\hat c$ are respectively the annihilation operators for the storage mode, auxiliary transmon, and resonator mode of bare angular frequencies $\omega_\mathrm{s0}$, $\omega_\mathrm{q0}$, and $\omega_\mathrm{r0}$. The auxiliary transmon is described as a Kerr oscillator with bare self-Kerr (anharmonicity) $K_{\mathrm{q}0}$. The auxiliary is coupled to both the storage and resonator modes through electric-dipole interactions described with Jaynes-Cummings coupling rates $g_\mathrm{sq}$ and $g_\mathrm{qr}$, respectively\,\cite{Blais2021}. The Purcell filter is neglected as it has negligible impact on the energy spectrum and is never populated during the experiments.
        
        Control of the system is performed via direct time-dependent driving of the storage and auxiliary modes and is captured by the generic drive terms
        \begin{align}
            \hat H^\mathrm{(JC)}_\mathrm{drive}(t)/\hbar&=\Omega_\mathrm{s}(t) \hat a^\dag+\Omega_\mathrm{q}(t)\hat b^\dag+\text{H.c.},
                \label{eq:driven_JC_Hamiltonian}
        \end{align}
        where $\Omega_\mathrm{s}(t)$ and $\Omega_\mathrm{q}(t)$ represent the time-dependent drive applied to the bare storage and auxiliary modes, respectively. No drive on the resonator is considered here given that the auxiliary readout process is not simulated.
    
        Given that all three modes are far detuned from one another, one can use a simplified model obtained from second-order perturbation theory\,\cite{Blais2021} in which the static part is described by the following dispersive Hamiltonian
        \begin{align}
            \hat H_\mathrm{disp}/\hbar&=\underbrace{\vphantom{\frac{K_{\mathrm{q}0}}{2}}\omega_\mathrm{s}\hat a^\dagger\hat a+\omega_\mathrm{q}\hat b^\dagger\hat b+\omega_\mathrm{r}\hat c^\dagger\hat c}_\text{dressed modes}+\underbrace{\frac{K_\mathrm{s}}{2}\hat a^{\dagger2}\hat a^2+\frac{K_\mathrm{q}}{2}\hat b^{\dagger2}\hat b^2+\frac{K_\mathrm{r}}{2}\hat c^{\dagger2}\hat c^2}_\text{dressed self-Kerr}+\underbrace{\vphantom{\frac{K_{\mathrm{q}0}}{2}}2\chi_\mathrm{sq}\hat a^\dagger\hat a\hat b^\dagger\hat b+2\chi_\mathrm{qr}\hat b^\dagger\hat b\hat c^\dagger\hat c}_\text{cross-Kerr} \nonumber\\
            &+\underbrace{2\chi_\mathrm{s^2q}\hat a^{\dagger2}\hat a^2\hat b^\dagger\hat b+2\chi_\mathrm{sq^2}\hat a^\dagger\hat a\hat b^{\dagger2}\hat b^2+2\chi_\mathrm{q^2r}\hat b^{\dagger2}\hat b^2\hat c^\dagger\hat c+2\chi_\mathrm{qr^2}\hat b^\dagger\hat b\hat c^{\dagger2}\hat c^2}_\text{second-order cross-Kerr}.
            \label{eq:dispersive_Hamiltonian}
        \end{align}
        In contrast to the bare modes in Eq.\,\eqref{eq:bare_Hamiltonian}, here the annihilation operators $\hat a$, $\hat b$, and $\hat c$ correspond to dressed modes, \textit{i.e.}, the auxiliary transmon is now slightly hybridized with the storage and resonator modes and vice versa. Higher-order cross- and single-mode interaction terms have been neglected as well as the storage-resonator cross-Kerr term $2\chi_\mathrm{sr}\hat a^\dagger\hat a\hat c^\dagger\hat c$.
    
        The driving Hamiltonian also needs to be consistently modified. Here, we go further by assuming specific forms of the driving amplitudes $\Omega_\mathrm{s,q}(t)$ and by neglecting off-resonant terms in a standard rotating wave approximation (RWA). Specifically, we assume that the angular frequency $\omega_\mathrm{d,s}$ of the drive $\Omega_\mathrm{s}(t)$ on the storage mode, described by
        \begin{align}
            \Omega_\mathrm{s}(t)=\epsilon_\mathrm{s}(t)\exp(-i\omega_\mathrm{d,s} t)+\text{c.c.},
        \end{align}
        is detuned by $\chi_\mathrm{sq}$ from resonance at $\omega_\mathrm{s}$, \textit{i.e.}~$\Delta_\mathrm{s}=\omega_\mathrm{d,s}-\omega_\mathrm{s}=\chi_\mathrm{sq}$ with a slowly varying envelope $\epsilon_\mathrm{s}(t)$, and away from the strong drive regime with $|\epsilon_\mathrm{s}(t)|/|\omega_\mathrm{s}-\omega_\mathrm{q}|, |\epsilon_\mathrm{s}(t)/\omega_\mathrm{s}|\ll1$ at all times $t$.

        The drives on the auxiliary transmon serves two main purposes and read
        \begin{align}
            \Omega_\mathrm{q}(t)=\epsilon_\mathrm{q}(t)e^{-i\omega_\mathrm{d,q} t}+\epsilon_{f0g1}(t)e^{-i\omega_{\mathrm{d},f0g1} t}+\text{c.c.}
        \end{align}
        The first drive has a frequency resonant with the auxiliary transmon dressed $|g\rangle\leftrightarrow|e\rangle$ or $|e\rangle\leftrightarrow|f\rangle$ transitions, \textit{i.e.}~$\omega_\mathrm{d,q}=\omega_\mathrm{q}$ or $\omega_\mathrm{d,q}= \omega_\mathrm{q}+K_\mathrm{q}$ respectively, with a slowly varying envelope $ \epsilon_\mathrm{q}(t)$. This drive allows one to implement auxiliary rotations and $|e\rangle\leftrightarrow|f\rangle$ coherent swaps. The second drive, at frequency $\omega_{\mathrm{d},f0g1}=\omega_{f0g1}\approx2\omega_\mathrm{q}+ K_\mathrm{q}-\omega_\mathrm{r}$, is resonant with the $|f0_\mathrm{r}\rangle \leftrightarrow |g1_\mathrm{r}\rangle$ transition and allows for the dissipative swaps to be performed during the auxiliary reset (see main text). The driving amplitudes are small enough to neglect all non-resonant terms given that $|\epsilon_\mathrm{q}(t)|,|\epsilon_{f0g1}(t)|\ll|\omega_\mathrm{s}-\omega_\mathrm{q}|, |\omega_\mathrm{r}-\omega_\mathrm{q}|,|\omega_\mathrm{q}|$ at all times $t$.

        The resulting driving Hamiltonian reads
        \begin{align}
            \hat H_\mathrm{drive}(t)/\hbar\approx\underbrace{\vphantom{\frac{1}{\sqrt{2}}}\delta\omega_\mathrm{Stark}(t)\hat b^\dag\hat b}_\text{ac Stark shift}+\underbrace{\vphantom{\frac{1}{\sqrt{2}}}\epsilon_\mathrm{q}(t)e^{-i\omega_\mathrm{d,q}t}\hat b^\dag+ \epsilon_\mathrm{s}(t)e^{-i\omega_\mathrm{d,s}t}\hat a^\dag+\text{H.c.}}_\text{RWA single-mode drives}+\underbrace{\frac{1}{\sqrt{2}}\tilde g(t)e^{-i\omega_{d,f0g1}t}\hat b^{\dag 2}\hat c+\text{H.c.}}_\text{$f0g1$ effective interaction},
                \label{eq:driven_dispersive_Hamiltonian}
        \end{align}
        where the ac Stark shift,
        \begin{align}
            \delta\omega_\mathrm{Stark}(t) = -\frac{2K_\mathrm{q0}(K_\mathrm{q0} +2\omega_\mathrm{q0} + 2\Delta_\mathrm{qr})}{\Delta_\mathrm{qr}(K_\mathrm{q0} + \Delta_\mathrm{qr})(K_\mathrm{q0} - \Delta_\mathrm{qr})}|\epsilon_{f0g1}(t)|^2,
            \label{eq:ac_stark_shift}
        \end{align}
        mainly results from the strong $\epsilon_{f0g1}(t)$ drive and where the effective $|f0_r\rangle \leftrightarrow |g1_r\rangle$ interaction is given by
        \begin{align}
            \tilde g(t) = -\sqrt{2}\frac{K_\mathrm{q0}g_\mathrm{qr}}{\Delta_\mathrm{qr}(K_\mathrm{q0}-\Delta_\mathrm{qr})}\epsilon_{f0g1}(t),
            \label{eq:gtilde}
        \end{align}
        where the detuning $\Delta_\mathrm{qr}=\omega_\mathrm{q0}-\omega_\mathrm{r0}$.

        In order to map adequately the above theoretical model onto the experiments, all relevant energy rates need to be characterized. The method used consists of first measuring experimentally the dressed frequencies $\omega_\mathrm{s}$, $\omega_\mathrm{q}$ and $\omega_\mathrm{r}$, their respective dispersive shifts due to the cross-Kerr coefficients $\chi_\mathrm{sq}$ (see Sec.\,\ref{ssec:initial_calibration}) and $\chi_\mathrm{qr}$, and the auxiliary dressed transmon anharmonicity $K_\mathrm{q}$. From the measured values, a numerical fit is used to reconstruct the associated Jaynes-Cumming Hamiltonian $\hat H_\mathrm{JC}$ and extract the bare frequencies $\omega_\mathrm{s0}$, $\omega_\mathrm{q0}$ and $\omega_\mathrm{r0}$, the bare Kerr nonlinearity $K_\mathrm{q0}$ and Jaynes-Cumming coupling coefficients $g_\mathrm{sq}$ and $g_\mathrm{qr}$. The numerically estimated values for the coupling coefficients are given in Table \ref{tab:Hamiltonian_parameters}. Once the Jaynes-Cumming Hamiltonian is reconstructed, its numerical diagonalization allows to estimate the rates in Eq.\,\eqref{eq:dispersive_Hamiltonian} that are not directly measured experimentally, such as the coefficients for the second-order cross-Kerr and induced self-Kerr. Table \ref{tab:Hamiltonian_parameters} gives most of the values resulting from this procedure and required for the simulations.
            
        For sanity checks, the numerically-estimated model is compared to analytical expressions derived in Ref.\,\cite{Wang2021a}, which are given by
        \begin{align}
            \omega_\mathrm{s,r}&=\omega_{\mathrm{s,r}0}-\frac{g_\mathrm{sq,qr}^2}{\Delta_\mathrm{sq,qr}},
            \label{eq:dressed_frequency_sr} \\
            \omega_\mathrm{q}&=\omega_{\mathrm{q},0}+\frac{g_\mathrm{sq}^2}{\Delta_\mathrm{sq}}+\frac{g_\mathrm{qr}^2}{\Delta_\mathrm{qr}},
            \label{eq:dressed_frequency_q} \\
            K_\mathrm{s,r}&=\frac{2g_\mathrm{sq,qr}^4K_{\mathrm{q}0}}{\Delta_\mathrm{sq,qr}\left(2\Delta_\mathrm{sq,qr}+K_{\mathrm{q}0}\right)},
            \label{eq:self-Kerr_sr} \\              
            K_\mathrm{q}&=K_{\mathrm{q}0}-\frac{2g_\mathrm{sq}^2K_{\mathrm{q}0}}{\Delta_\mathrm{sq}\left(\Delta_\mathrm{sq}+K_{\mathrm{q}0}\right)}-\frac{2g_\mathrm{qr}^2K_{\mathrm{q}0}}{\Delta_\mathrm{qr}\left(\Delta_\mathrm{qr}+K_{\mathrm{q}0}\right)},
            \label{eq:self-Kerr_q} \\                
            \chi_\mathrm{sq,qr}&=\frac{g_\mathrm{sq,qr}^2K_{\mathrm{q}0}}{\Delta_\mathrm{sq,qr}\left(\Delta_\mathrm{sq,qr}+K_{\mathrm{q}0}\right)},
            \label{eq:cross-Kerr} \\                
            \chi_\mathrm{s^2q,qr^2}&=\frac{-g_\mathrm{sq,qr}^4K_{\mathrm{q}0}^2\left(3K_{\mathrm{q}0}^3+11K_{\mathrm{q}0}^2\Delta_\mathrm{sq,qr}+15K_{\mathrm{q}0}\Delta_\mathrm{sq,qr}^2+9\Delta_\mathrm{sq,qr}^3\right)}{\Delta_\mathrm{sq,qr}^3\left(2\Delta_\mathrm{sq,qr}+K_{\mathrm{q}0}\right)\left(2\Delta_\mathrm{sq,qr}+3K_{\mathrm{q}0}\right)\left(\Delta_\mathrm{sq,qr}+K_{\mathrm{q}0}\right)^3},
            \label{eq:nonlinear_cross-Kerr} \\
            \chi_\mathrm{sr}&=\frac{g_\mathrm{sq}^2g_\mathrm{qr}^2K_\mathrm{q0}\left(\Delta_\mathrm{sq}+\Delta_\mathrm{qr}\right)}{\Delta_\mathrm{sq}^2\Delta_\mathrm{qr}^2\left(\Delta_\mathrm{sq}+\Delta_\mathrm{qr}+K_\mathrm{q0}\right)},
            \label{eq:cross-Kerr_sr}
        \end{align}
        where the detuning $\Delta_\mathrm{sq}=\omega_\mathrm{q0}-\omega_\mathrm{s0}$. The analytical values and the numerically estimations are compared in Table \ref{tab:Hamiltonian_parameters} and shows excellent agreement.
            
        \begin{table*}[t]
            \centering
            \begin{tabular}{l|l|l||l|l|l|r||l|l}
                Parameter & Mode(s) & Symbol & Experimental & Numerical & Analytical & Units & Eq. & Comments\\
                \hline\hline
                Self-Kerr & Storage & $K_\mathrm{s}/2\pi$ & --- & $-0.335$ & $-0.334$ & Hz & \eqref{eq:self-Kerr_sr} & \\
                & Auxiliary & $K_\mathrm{q}/2\pi$ & $-0.257$ & $-0.257$ & $-0.257$ & GHz & \eqref{eq:self-Kerr_q} & \textit{Fit input}\\
                & Resonator & $K_\mathrm{r}/2\pi$ & --- & $-0.772$ & $-0.781$ & kHz & \eqref{eq:self-Kerr_sr} & \\
                \hline
                Cross-Kerr & Storage-auxiliary & $\chi_\mathrm{sq}/2\pi$ & $-10.72$ & $-10.72$ & $-10.72$ & kHz & \eqref{eq:cross-Kerr} & \textit{Fit input}, see Fig.\,\ref{fig:displacement_frequency_calibration}\\
                & Auxiliary-resonator & $\chi_\mathrm{qr}/2\pi$ & $-0.392$ & $-0.392$ & $-0.394$ & MHz & \eqref{eq:cross-Kerr} & \textit{Fit input}\\
                & Storage-resonator & $\chi_\mathrm{sr}/2\pi$ & --- & $+21.2$ & $+23.6$ & Hz & \eqref{eq:cross-Kerr_sr} & Neglected in sim.\\
                \hline
                Second-order & Storage-auxiliary & $\chi_\mathrm{s^2q}/2\pi$ & --- & $-0.230$ & $-0.230$ & Hz & \eqref{eq:nonlinear_cross-Kerr} & \\
                cross-Kerr & Storage-auxiliary & $\chi_\mathrm{sq^2}/2\pi$ & --- & $-4.232$ & --- & kHz & & Used $-12.2$\,kHz in sim.\\
                & Auxiliary-resonator & $\chi_\mathrm{qr^2}/2\pi$ & --- & $+0.207$ & $+0.211$ & kHz & \eqref{eq:nonlinear_cross-Kerr} & \\
            \end{tabular}
            \caption{
            \textbf{Parameters characterizing the coherent dynamics.}
            Parameters labeled as \textit{Fit input} are used to numerically reconstruct the Jaynes-Cummings Hamiltonian of Eq.\,\eqref{eq:bare_Hamiltonian} from a fitting procedure. When applicable, the equation for the analytical calculation are referenced.}
            \label{tab:Hamiltonian_parameters}
        \end{table*}

    \subsection{Dissipative dynamics and Lindbladian parameters}

        In addition to the coherent dynamics effectively described by the dispersive Hamiltonian of Eq.\,\eqref{eq:dispersive_Hamiltonian} and its driving counterpart of Eq.\,\eqref{eq:driven_dispersive_Hamiltonian}, the system composed of the storage, auxiliary and resonator modes undergo a non-unitary evolution induced by the coupling to the environment. The model used here to capture this dissipative dynamics assumes an independent Markovian bath for each of these three modes so that the resulting non-unitary time evolution is described by a standard Lindblad master equation of the form
        \begin{align}
            \partial_t \hat\rho(t)=-\frac{i}{\hbar}[\hat H_\mathrm{disp}+\hat H_\mathrm{drive}(t),\hat \rho(t)]+\sum_j \frac{\kappa_j}{2}\left(2\hat O_j \hat\rho(t)\hat O^\dag_j-\hat O^\dag_j\hat O_j \hat\rho(t)-\hat\rho(t)\hat O^\dag_j\hat O_j\right).
            \label{eq:Lindblad}
        \end{align}
        Here, $\hat\rho(t)$ is the density matrix describing the three-mode system and the sum over $j$ runs over all dissipative processes, each characterized by a jump operator $\hat O_j$ and a rate $\kappa_j$. Each mode $\mathrm{k}\in\{\mathrm{s,q,r}\}$ undergoes a decay process with rate $(1+\bar n^\mathrm{eq}_\mathrm{k})\kappa_{1\mathrm{k}}$ and respective jump operators $\hat O_\mathrm{k}\in\{\hat a,\hat b,\hat c\}$, an heating process with rate $\bar n^\mathrm{eq}_\mathrm{k}\kappa_{1\mathrm{k}}$ and respective jump operators $\hat O_\mathrm{k}\in\{\hat a^\dag,\hat b^\dag,\hat c^\dag\}$, and finally a dephasing process with rate $2\kappa_{\phi\mathrm{k}}$ and respective jump operators $\hat O_\mathrm{k}\in\{\hat a^\dag\hat a,\hat b^\dag\hat b,\hat c^\dag\hat c\}$. The equilibrium population of mode $\mathrm{k}$ is given by $n^\mathrm{eq}_\mathrm{k}$. A summary of all dissipative rates is presented in Table \ref{tab:Lindbladian_parameters}. 

        \begin{table*}[t]
            \centering
            \begin{tabular}{l|l||l|l|r||l}
                Parameter & Mode or state & Symbol & Value & Units & Comment\\
                \hline\hline
                Decay rate & Storage & $\kappa_{1\mathrm{s}}/2\pi$ & $0.474$ & kHz & See Fig.\,\ref{fig:storage_relaxation}\\
                & Auxiliary & $\kappa_{1\mathrm{q}}/2\pi$ & $4.81$ & kHz & See Fig.\,\ref{fig:auxiliary_coherence}\\
                & Resonator & $\kappa_{1\mathrm{r}}/2\pi$ & $1.73$ & MHz &\\
                \hline
                Dephasing rate & Storage & $\kappa_{\phi\mathrm{s}}/2\pi$ & $1.45$ & Hz & See Sec.\,\ref{sssec:init_error_budget}\\
                & Auxiliary & $\kappa_{\phi\mathrm{q}}/2\pi$ & $0.904$ & kHz & See Fig.\,\ref{fig:auxiliary_coherence}\\
                & Resonator & $\kappa_{\phi\mathrm{r}}/2\pi$ & $0$ & --- & Assumed, not measured\\
                \hline
                Equilibrium population & Storage & $\bar n_\mathrm{s}^\mathrm{eq}$ & $0$ & --- & Assumed, not measured\\
                & Auxiliary & $\bar n_\mathrm{q}^\mathrm{eq}$ & $0.38$ & $\%$ &\\
                & Resonator & $\bar n_\mathrm{r}^\mathrm{eq}$ & $0$ & --- & Assumed, not measured\\
                \hline
                Readout error & Auxiliary ground state & $\varepsilon_g$ & 0.12 & \% & See Sec.\,\ref{ssec:readout}\\
                & Auxiliary excited state & $\varepsilon_e$ & 0.45 & \% & See Sec.\,\ref{ssec:readout}\\
            \end{tabular}
            \caption{
            \textbf{Parameters characterizing the dissipative dynamics.}
            When applicable, the figure or section discussing how the parameters are obtained is referenced.}
            \label{tab:Lindbladian_parameters}
        \end{table*}

    \subsection{Dynamics in the displaced and rotating storage frame}

        During the GKP states initialization and quantum error correction, a large number of photons is coherently injected in the storage mode as a control drive $\epsilon_\mathrm{s}(t)$ is applied [Eq.\,\eqref{eq:driven_dispersive_Hamiltonian}]. In order to perform efficient simulations even when the storage mode is largely populated, one needs to work in a displaced and rotating time-dependent storage mode frame. Mathematically, this means applying the unitary transformation $\hat U(t)=\hat U_\mathrm{disp}(t)\hat U_\mathrm{rot}(t)$, with $\hat U_\mathrm{disp}(t) =\exp\left(\alpha^*(t)\hat a-\alpha(t)\hat a^\dag\right)$ and $\hat U_\mathrm{rot}(t)=\exp(i\omega_\mathrm{d,s}t)$, such that
        \begin{align}
            \partial_t\hat\rho_\mathrm{DR}(t)=-\frac{i}{\hbar}[\hat H_\mathrm{DR}(t),\hat\rho_\mathrm{DR}(t)]+\sum_j\frac{\kappa_j}{2}\left(2\hat O_j \hat\rho_\mathrm{DR}(t)\hat O^\dag_j-\hat O^\dag_j\hat O_j \hat\rho_\mathrm{DR}(t)-\hat\rho_\mathrm{DR}(t)\hat O^\dag_j\hat O_j\right)+\mathcal{D}_\mathrm{disp}[\hat\rho_\mathrm{DR}(t)],
            \label{eq:displaced_Lindblad}
        \end{align}
        where the state in this frame, $\hat\rho_\mathrm{DR}(t)=\hat U(t)\rho(t)\hat U(t)^\dag$, evolves under the transformed Hamiltonian
        \begin{align}
            \hat H_\mathrm{DR}(t)&=\hat U(t)\left(\hat H_\mathrm{disp}+\hat H_\mathrm{drive}(t)\right)\hat U(t)^\dag+i\{\partial_t \hat U(t)\}\hat U^\dag(t)-i\frac{\kappa_{1\mathrm{s}}}{2}\left(\alpha(t)\hat a^\dag-\alpha^*(t)\hat a\right),\nonumber\\
            &=\hat H_\mathrm{disp}^\mathrm{DR}(t)+\hat H_\mathrm{drive}^\mathrm{DR}(t),
            \label{eq:displaced_Hamiltonian}
        \end{align}
        and is affected by additional dissipation terms coming from the storage mode displacement that are captured by $\mathcal{D}_\mathrm{disp}[\hat\rho_\mathrm{DR}(t)]$.
        
        The explicit form of $\hat H_\mathrm{disp}^\mathrm{DR}(t)$ is obtained from Eq.\,\eqref{eq:dispersive_Hamiltonian} by performing the substitutions 
        \begin{align}
            \hat a&\rightarrow\hat a+\alpha(t),\nonumber\\
            \hat a^\dag&\rightarrow\hat a^\dag+\alpha^*(t),\\
            \omega_\mathrm{s}&\rightarrow-\Delta_\mathrm{s}=\omega_\mathrm{s}-\omega_\mathrm{d,s},\nonumber 
        \end{align}
        where $\Delta_\mathrm{s}$ is the storage mode drive detuning.
            
        The transformed driving Hamiltonian $\hat H_\mathrm{drive}^\mathrm{DR}(t)$ is obtained from Eq.\,\eqref{eq:driven_dispersive_Hamiltonian} and the following substitution of the storage mode drive amplitude 
        \begin{align}
            \epsilon_\mathrm{s}(t)e^{-i\omega_\mathrm{d,s}t} \rightarrow \epsilon_\mathrm{s}(t) - i\partial_t \alpha(t) - i\frac{\kappa_{1\mathrm{s}}}{2}\alpha(t). 
        \end{align}
        Finally, the additional dissipation processes that explicitly appears in the displaced frame reads
        \begin{align}
            \mathcal{D}_\mathrm{disp}[\hat\rho_\mathrm{DR}(t)]&=\kappa_{\phi\mathrm{s}}\left(2\hat A^\dag\hat A\hat\rho_\mathrm{DR}(t)\hat A^\dag\hat A-\hat A^\dag\hat A\hat A^\dag\hat A\hat\rho_\mathrm{DR}(t)-\hat\rho_\mathrm{DR}(t)\hat A^\dag\hat A\hat A^\dag\hat A\right)\nonumber\\
            &-\kappa_{\phi \mathrm{s}}\left(2\hat a^\dag\hat a\hat\rho_\mathrm{DR}(t)\hat a^\dag\hat a-\hat a^\dag\hat a\hat a^\dag \hat a\hat\rho_\mathrm{DR}(t)-\hat\rho_\mathrm{DR}(t)\hat a^\dag\hat a\hat a^\dag \hat a\right),
        \end{align}
        with the time-dependent jump operator $\hat A=\hat a+\alpha(t)$.
    
        At this point, the displacement transformation $\hat U_\mathrm{disp}(t)$ is completely general. In order to make the numerical simulations efficient, the goal is to suppress the term linear in $\hat a$ with a strategic choice of $\alpha(t)$ such that the required Hilbert space does not increase in size as the storage mode drive is applied. In this work, 
        \begin{align}
            \alpha(t) = -i\int_0^t \mathrm{d}\tau\ \epsilon_\mathrm{s}(\tau),
        \end{align}
        is chosen to suppress the largest contribution directly proportional to the drive amplitude while keeping the expression simple and independent of the storage mode decay and dephasing rates. This choice is motivated by the fact that $\left|\kappa_{1\mathrm{s}}\alpha(t)\right|,\left|\kappa_{\phi\mathrm{s}}\alpha(t)^2\alpha^*(t)\right|\ll\left|\Delta_\mathrm{s}\right|$, leading to negligible effective displacements caused by the dissipation channels.
            
    \subsection{Definition of pulse parameters}

        The pulse envelopes $\epsilon_\mathrm{s}(t)$, $\epsilon_\mathrm{q}(t)$, and $\epsilon_{f0g1}(t)$ used in the experiments can all be described as Gaussian pulses of width $\Theta$ with an optional plateau of duration $\tau$. The width $\Theta$ is defined such that the integral of the pulse is the same as for a square pulse with a plateau $\tau=\Theta$. More explicitly, the generic pulse envelope is given by
        \begin{align}
            \epsilon(t)=\epsilon_0\times\left\{
            \begin{array}{lr}
            e^{-\pi\left(t-t_0\right)^2/\Theta^2} & \mbox{if } t<t_0-\tau/2,\\
            1 & \mbox{if } t_0-\tau/2\leq t<t_0+\tau/2,\\
            e^{-\pi\left(t-t_0\right)^2/\Theta^2} & \mbox{if } t\geq t_0+\tau/2,\\
            \end{array}
            \right.
            \end{align}
         where $t_0$ is the time $t$ at which $\epsilon(t)=\epsilon_0$. The spacing $\delta t_\mathrm{ab}$ between pulses $\mathrm{a}$ and $\mathrm{b}$ is defined as $\delta t_\mathrm{ab}=\left(t_\mathrm{b}-T_\mathrm{b}/2-\tau_\mathrm{b}/2\right)-\left(t_\mathrm{a}+T_\mathrm{a}/2+\tau_\mathrm{a}/2\right)$.
        
\section{Auxiliary transmon}

    \subsection{Relaxation and coherence}

\begin{figure*}[t]
            \centering
            \includegraphics[scale=0.81]{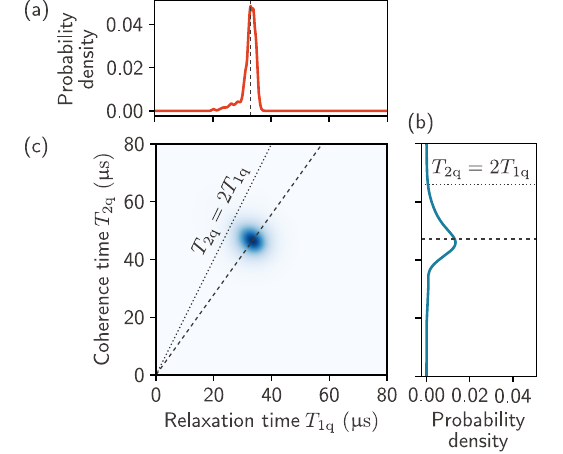}
            \caption{
            \textbf{Relaxation and coherence of the auxiliary.}
            Probability density for the (a)~relaxation time $T_{1\mathrm{q}}$ and (b)~echo coherence time $T_{2\mathrm{q}}$ of the auxiliary. The dashed lines indicate the respective median of $T_{1\mathrm{q}}=33\,\unit{\micro\second}$ and $T_{2\mathrm{q}}=47\,\unit{\micro\second}$.
            (c)~Joint probability density of $T_{1\mathrm{q}}$ and $T_{2\mathrm{q}}$ calculated using pairs of subsequent measurements. The dashed line indicates a linear fit.
            In (b) and (c), the dotted line indicates the $T_1$ limit $T_{2\mathrm{q}}=2T_{1\mathrm{q}}$.
            }
            \label{fig:auxiliary_coherence}
        \end{figure*}

         Figure~\ref{fig:auxiliary_coherence}(a--b) shows the probability densities of the auxiliary relaxation time $T_{1\mathrm{q}}$ and echo coherence time $T_{2\mathrm{q}}$ from 259 pairs of measurements scattered across about 35 days. The probability densities are calculated assuming a Gaussian distribution centered at each value and with a standard deviation given by the fitting error on each value. The joint probability density, shown in Fig.\,\ref{fig:auxiliary_coherence}(c), is calculated using pairs of subsequent measurements.

         The echo coherence time $T_{2\mathrm{q}}$ is lower than the $T_1$ limit of $T_{2\mathrm{q}}=2T_{1\mathrm{q}}$, indicating the presence of pure dephasing at a rate $\kappa_{\phi\mathrm{q}}=\left(1/T_{2\mathrm{q}}\right)-\kappa_{1\mathrm{q}}/2$, where $\kappa_{1\mathrm{q}}=1/T_{1\mathrm{q}}$ is the auxiliary relaxation rate. The echo coherence time is the relevant coherence metric given that the initialization, quantum error correction, and tomography protocols intrinsically use dynamical decoupling through echoed conditional displacements\,\cite{Eickbusch2022}.

    \subsection{Rotations}

        Auxiliary rotations with a rotation axis in the equatorial plane (so-called \textit{XY rotations}), fully parameterized with a complex number $R$, correspond to the gate
        \begin{align}
            \hat U_R=\mathrm{exp}\left(-i\left[\hat\sigma_x\mathrm{Re}(R)+\hat\sigma_y\mathrm{Im}(R)\right]/2\right),
        \end{align}
        where $\boldsymbol{\hat\sigma}=\left(\hat\sigma_x,\hat\sigma_y,\hat\sigma_z\right)$ are the Pauli operators of the manifold of the first two states of the auxiliary, $\ket{g}$ and $\ket{e}$. The DRAG protocol is implemented to suppress detuning errors caused by the presence of the auxiliary second excited state $\ket{f}$\,\cite{Motzoi2009}. In this protocol, the pulse shape $\epsilon_\mathrm{q}(t)$ used to implement auxiliary rotations is changed to
        \begin{align}
            \epsilon_\mathrm{q}(t)\rightarrow\epsilon_\mathrm{q}(t)+i\tau_\mathrm{D}\dot\epsilon_\mathrm{q}(t),
        \end{align}
        where $\tau_\mathrm{D}$ is the DRAG coefficient. Rotations around the $\hat\sigma_z$ axis (so-called \textit{Z rotations}), parameterized by the phase $\varphi_z$, correspond to the gate
        \begin{align}
            \hat U_{\varphi_z}=\mathrm{exp}\left(-i\hat\sigma_z\varphi_z/2\right).
        \end{align}
        These rotations around the $\hat\sigma_z$ axis are implemented with so-called virtual-Z gates by updating the phase reference of subsequent rotations\,\cite{McKay2017a}. Using an error amplification scheme, closed-loop optimization of the drive amplitude required for a rotation of amplitude $|R|=\pi/2$ and of the DRAG coefficient $\tau_\mathrm{D}$ is used to minimize coherent errors.
        
    \subsection{Readout}
    \label{ssec:readout}

        \subsubsection{Excited state promoted readout}

            As previously mentioned, a lumped-element Josephson parametric amplifier (LJPA) is present in the readout line, but is not used. The bias current is set to minimize the absorption of the readout signal when the LJPA is not pumped.
            
            The autonomous QEC scheme used here only requires the auxiliary to be read out at the end of the tomography protocol, \textit{i.e.} an end-of-the-line measurement. The auxiliary readout therefore does not need to be a quantum nondemolition measurement. With this consideration, the excited state promotion technique\,\cite{Mallet2009} is used to decrease the readout errors. The excited state promotion readout prepends the readout pulse with a $\pi$ pulse on the $\ket{e}\leftrightarrow\ket{f}$ transition. If the auxiliary is in the ground state, this pulse, ideally, does nothing. If, however, the auxiliary is in the first excited state $\ket{e}$, the excited state promotion pulse sends it to the second excited state $\ket{f}$.
            
            This has two benefits. First, the dispersive shift between $\ket{g}$ and $\ket{f}$ is larger than between $\ket{g}$ and $\ket{e}$ for the frequency configuration of the device, leading to a larger separation in phase space for the same readout amplitude and duration provided the readout frequency is appropriately chosen. Secondly, if a decay event happens during the readout, leading to $\ket{f}\rightarrow\ket{e}$, the readout result can be unaffected provided than the first excited state is well separated from the ground state, which can be achieved, for example, by an optimization of the readout frequency.
            
            The excited state promoted readout is used for the experimental results presented in the figures of the main text, except for measurements on the average reset error of Fig.\,2(c). In that case, it is desired to keep the distinction between the first and second excited states, and not to interchange them.

        \subsubsection{Readout visibility and state discrimination}

            Experimentally, the single-shot readout of the auxiliary is benchmarked by evaluating the readout visibility, defined as
            \begin{align}
                \mathcal{V}=p_{g|g}-p_{g|e},
            \end{align}
            where $p_{g|k}$ is the probability of assigning the readout result to the ground state $\ket{g}$ when imperfectly preparing, prior to the readout pulses (which include the $\pi$ pulse addressing the $\ket{e}\leftrightarrow\ket{f}$ transition), the auxiliary state $\ket{k}$ with $k\in\left\{g,e\right\}$. The readout visibility therefore includes initialization and control errors.
                
            The readout visibility is obtained as follows. The raw signal on the digitizer is demodulated with a rectangular demodulation window. The two-dimensional histogram of the demodulated signal in the complex plane is computed and fitted to a two-dimensional Gaussian in order to find the complex signal amplitudes $V_k$ corresponding to state $\ket{k}$. Following this calibration, each single-shot raw measurement result $V_n$, where $n\in\left\{0,1,2\dots,N-1\right\}$, where $N$ is the number of shots, can be assigned a state based on the following algorithm: $V_n$ is assigned to $\ket{k}$ if $\left|V_n-V_k\right|<w\left|V_e-V_g\right|$. Otherwise, it is excluded. The dimensionless parameter $w$ describes the distance in the complex plane of the exclusion region relative to the distance between $V_g$ and $V_e$. For $w=0$, no shots are excluded.

\begin{figure*}[t]
                \centering
                \includegraphics[scale=0.81]{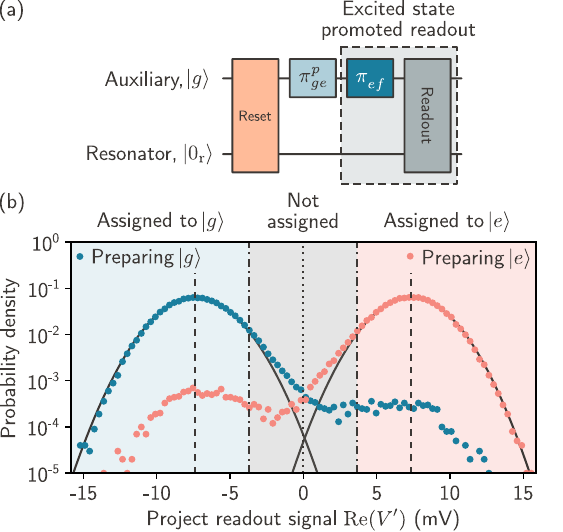}
                \caption{
                \textbf{Single-shot readout of the auxiliary.}
                (a)~Protocol for evaluating and optimizing the single-shot readout visibility and fidelity. The auxiliary is prepared in either the ground state $\ket{g}$ ($p=0$) or the excited state $\ket{e}$ ($p=1$) through the $\pi_{ge}^p$ pulse. A pre-reset prepends the auxiliary state preparation to lower the initial auxiliary excited state probability. The excited state promoted readout prepends the usual readout pulse with a $\pi_{ef}$ pulse.
                (b)~Histogram of the projected readout signal $\mathrm{Re}(V')$ for $N=10^5$ shots when imperfectly preparing the auxiliary ground and excited states $\ket{g}$ (blue) and $\ket{e}$ (pink), respectively. Solid lines are Gaussian fits to highlight the readout signals corresponding to $\mathrm{Re}(V_g')$ and $\mathrm{Re}(V_e')$ (vertical dashed lines). The threshold, at $\mathrm{Re}(V')=0$ by definition, is shown by the dotted line. The dot-dashed lines delimit the exclusion region, here with $w=0.5$, in which shots are excluded.
                }
                \label{fig:auxiliary_single_shot_readout}
            \end{figure*}

            Figure~\ref{fig:auxiliary_single_shot_readout} shows an example of a histogram of projected readout signal when preparing, to the best of our ability, both the ground state $\ket{g}$ and excited state $\ket{e}$. A translation and rotation in the complex plan, changing $V\rightarrow V'$, is used such that $\mathrm{Re}\left(V_g'\right)=-\mathrm{Re}\left(V_e'\right)$ and $\mathrm{Im}\left(V_k'\right)=0$. This enables one to present a one-dimensional histogram of $\mathrm{Re}\left(V'\right)$. Note that this is not necessary to assign a state to the raw readout results $V_n$. In the example shown in Fig.\,\ref{fig:auxiliary_single_shot_readout}, $1-p_{g|g}=0.87\%$ and $p_{g|e}=1.24\%$ for $w=0$, corresponding to a readout visibility $\mathcal{V}=0.9788$. Increasing $w$ to 0.5, $1-p_{g|g}=0.57\%$ and $p_{g|e}=1.01\%$ while throwing away less than $5\%$ of the shots. The readout visibility is improved to $\mathcal{V}=0.9842$.       
            
            Readout visibility measurements were interleaved during measurements of the logical lifetime, the main results of the paper. Averaging all of these results, we get $1-p_{g|g}=0.50\%$ and $p_{g|e}=0.83\%$ for $w=0.5$, corresponding to $\mathcal{V}=0.9867$. A rejection region parameter $w=0.5$ is used in the experimental results presented in the figures of the main text, except for measurements on the average reset error of Fig.\,2(c), in which case averaged measurements are used as discussed in Sec.\,\ref{ssec:reset}.

        \subsubsection{Estimation of readout errors}
        \label{ssec:readout_errors}

            The readout errors $\varepsilon_g$ and $\varepsilon_e$ are respectively defined as the probabilities $1-p_{g|g}$ and $p_{g|e}$ in the absence of initialization and control errors. In other words, $\varepsilon_g$ is the probability of \textit{not} assigning the ground state $\ket{g}$ when perfectly preparing $\ket{g}$, while $\varepsilon_e$ is the probability of assigning the ground state $\ket{g}$ when perfectly preparing $\ket{e}$. The readout fidelity is then simply given by $F_\mathrm{readout}=1-\varepsilon_g-\varepsilon_e$.
            
            Assuming that control errors are much smaller than both initialization and readout errors, the readout errors $\varepsilon_g$ and $\varepsilon_e$ are given by
            \begin{align}
                \label{eq:readout_errors}
                \varepsilon_g&=\frac{1-\left(2-p_{g|e}\right)p_e^\mathrm{eq}-p_{g|g}\left(1-p_e^\mathrm{eq}\right)}{1-2p_e^\mathrm{eq}},\\
                \varepsilon_e&=\frac{p_{g|e}\left(1-p_e^\mathrm{eq}\right)-p_{g|g}p_e^\mathrm{eq}}{1-2p_e^\mathrm{eq}}.\nonumber
            \end{align}
            Equations\,\eqref{eq:readout_errors} are used to estimate the readout errors $\varepsilon_g$ and $\varepsilon_e$ based on the measurements of $p_{g|g}$, $p_{g|e}$ and $p_e^\mathrm{eq}$ later discussed.
            
            For the values from interleaved measurements during the logical lifetime measurements, we get $\varepsilon_g=0.12\%$ and $\varepsilon_e=0.45\%$, corresponding to a readout fidelity $F_\mathrm{readout}=0.9942$. These readout errors, cited in Tab.\,\ref{tab:Lindbladian_parameters}, are used to correct the simulation results to include measurement errors given that the readout process is not simulated.

        \subsubsection{Closed-loop optimization}
        
            The optimization parameters for the auxiliary readout are the readout frequency, amplitude, and duration. The readout duration is given by the duration of the plateau given that a simple square readout pulse is used. The cost function $\mathcal{C}$ for the optimization is $\mathcal{C}=1-\mathcal{V}$. Indeed, even though the visibility include state preparation errors, these are independent on the readout optimization parameters.
            
            The optimizer used is the Bayesian optimization function \texttt{gp\_minimize} from the module \texttt{skopt}. The results presented in Fig.\,\ref{fig:auxiliary_single_shot_readout} are a check $6$ days after an optimization with $100$ iterations in total, among which one is an initial guess from a previous optimization and $19$ are random to sample the parameter space. The optimized value for the readout plateau is $1.06\,\unit{\micro\second}$.

    \subsection{Auxiliary excited state equilibrium population}
    \label{sec:thermal_population}

\begin{figure*}[t]
            \centering
            \includegraphics[scale=0.81]{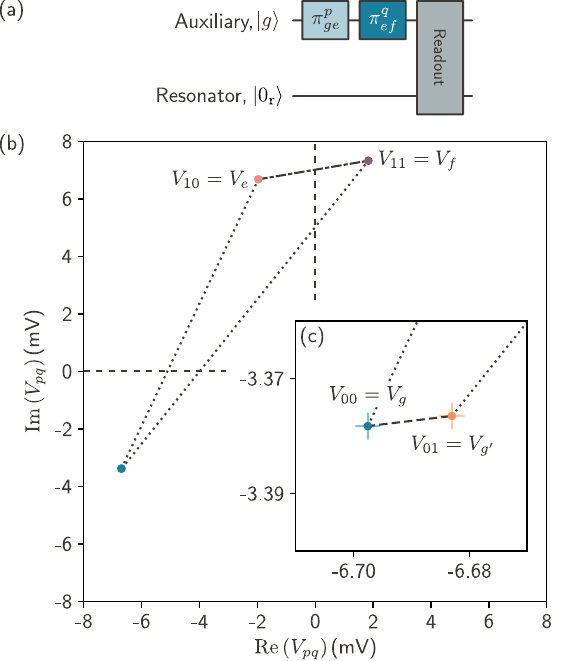}
            \caption{
            \textbf{Equilibrium population of the auxiliary.}
            (a)~Protocol to measure the auxiliary equilibrium excited state probability $p_e^\mathrm{eq}$. The auxiliary rotations $\pi_{ge}^p$ and $\pi_{ef}^q$ condition the prepared auxiliary state depending on $p$ and $q$.
            (b)~Raw demodulation voltages $V_{pq}$ averaged on $5\times10^6$ shots.
            (c)~Zoom-in to show the difference between $V_{00}$ and $V_{01}$, from which the equilibrium excited state population, here of $p_e^\mathrm{eq}=0.38(8)\%$, is proportional. Error bars are the $95\%$ confidence interval on the averaged values.
            }
            \label{fig:auxiliary_thermal_population}
        \end{figure*}

        The equilibrium population of the auxiliary excited state is measured with a method inspired by the so-called \textit{RPM method} of Ref.\,\cite{Geerlings2013}. The experiment consists in measuring the readout signals \textbf{without the excited state promotion pulse} when preparing either the ground state $\ket{g}$, the first excited state $\ket{e}$, the second excited state $\ket{f}$, or swapping the populations in $\ket{e}$ and $\ket{f}$ when preparing the ground state $\ket{g}$. The protocol is shown in Fig.\,\ref{fig:auxiliary_thermal_population}(a).

        More formally, the readout signals corresponding to each state preparation are $V_{pq}$ with $p,q\in\left\{0,1\right\}$, where a $\pi$-pulse on the $\ket{g}\leftrightarrow\ket{e}$ transition, $\pi_{ge}^p$, is applied if $p=1$. In a similar manner, a subsequent $\pi$-pulse on the $\ket{e}\leftrightarrow\ket{f}$ transition, $\pi_{ef}^q$, is applied if $q=1$. With this formalism, the ground state $\ket{g}$ is prepared with $p=0$ and $q=0$, the first excited state $\ket{e}$ with $p=1$ and $q=0$, the second excited state $\ket{f}$ with $p=1$ and $q=1$. Finally, the populations in $\ket{e}$ and $\ket{f}$ when preparing the ground state $\ket{g}$ is performed with $p=0$ and $q=1$. The same state preparation will be used for benchmarking the auxiliary reset errors in Sec.\,\ref{ssec:reset}.
    
        The auxiliary excited state probability can be calculated from the raw complex-valued demodulated readout signals with
        \begin{align}
            p_e^\mathrm{eq}=\frac{\left|V_{01}-V_{00}\right|}{\left|V_{01}-V_{00}\right|+\left|V_{10}-V_{11}\right|}.
        \end{align}
        The equilibrium population $\bar n_\mathrm{q}^\mathrm{eq}$ is related to the equilibrium excited state probability $p_e^\mathrm{eq}$ with
        \begin{align}
            \bar n_\mathrm{q}^\mathrm{eq}=\frac{p_e^\mathrm{eq}}{1-2p_e^\mathrm{eq}}.
        \end{align}
        Figure~\ref{fig:auxiliary_thermal_population}(b) and (c) show the measurements from which the value of $\bar n_\mathrm{q}^\mathrm{eq}=0.38(8)\%$ in Tab.\,\ref{tab:Lindbladian_parameters} comes from, where the demodulated readout signals $V_{pq}$ are averaged on $5\times10^6$ shots.
            
        The observed equilibrium population is equivalent to an effective temperature of $T_\mathrm{q}^\mathrm{eff}=64$\,mK calculated with $p_e^\mathrm{eq}=\exp\left(-\hbar\omega_\mathrm{q}/k_\mathrm{B}T_\mathrm{q}^\mathrm{eff}\right)$, where $k_\mathrm{B}$ is the Boltzmann constant.
            
        Interleaved measurements during the measurements of the logical lifetime give an average value of $0.41(33)\%$, where the error bar is the $95\%$ confidence interval calculated from the standard deviation across the multiple measurements each containing $10^5$ shots. The average value is consistent with the previously stated value.

    \subsection{Reset}
    \label{ssec:reset}

        \subsubsection{Estimating probabilities in averaged measurements}
        \label{ssec:probabilities_through_transformation}

\begin{figure*}[t]
                \centering
                \includegraphics[scale=0.81]{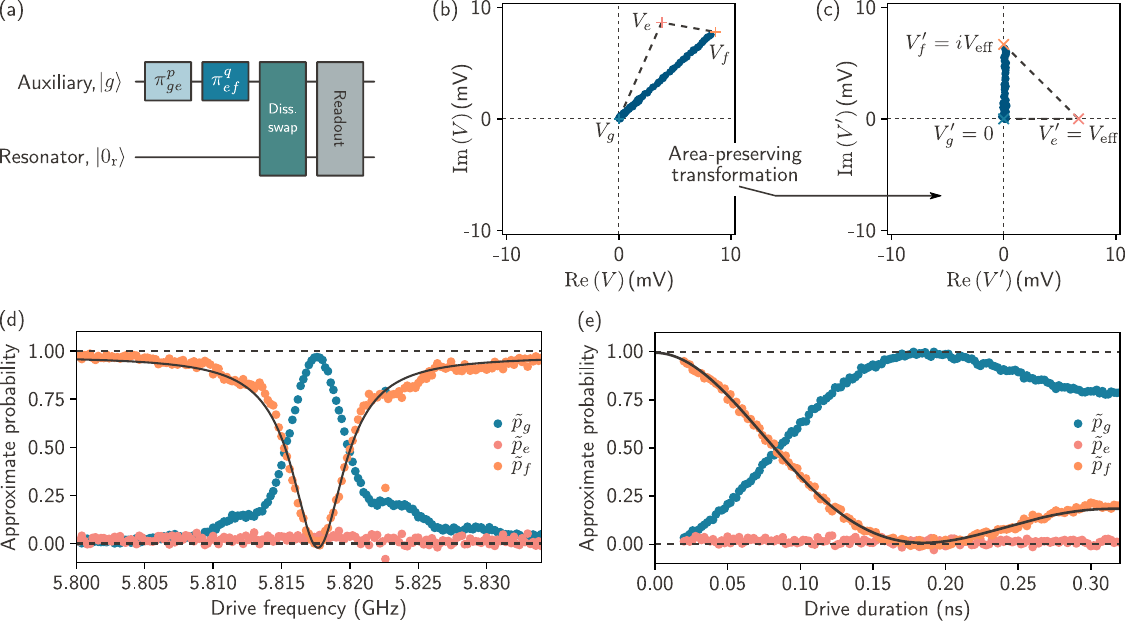}
                \caption{
                \textbf{Calibration of the auxiliary reset.}
                (a)~Protocol used to calibrate the auxiliary reset. The auxiliary is prepared in either of the ground state $\ket{g}$ ($p=0$, $q=0$), excited state $\ket{e}$ ($p=1$, $q=0$), or second excited state $\ket{f}$ ($p=1$, $q=1$), followed by a drive attempting to perform the $\ket{f0_\mathrm{r}}\leftrightarrow\ket{g1_\mathrm{r}}$ coherent swap. This coherent swap becomes the desired dissipative swap $\ket{f0_\mathrm{r}}\rightarrow\ket{g0_\mathrm{r}}$ through resonator decay. 
                (b)~Raw demodulation voltages $V_g$ (blue), $V_e$ (pink), and $V_f$ (orange).
                (c)~Demodulation voltages after the area-preserving transformation with $V_g'=0$, $V_e=V_\mathrm{eff}$, and $V_f'=iV_\mathrm{ef}$.
                The raw and corrected $V$ and $V'$ from which the results of Fig.\,\ref{fig:auxiliary_reset_calibration}(e) are obtained are shown in dark blue in (b) and (c), respectively.
                (d)~Approximate probability $\tilde p_g$ (blue), $\tilde p_e$ (pink), and $\tilde p_f$ (orange) of the auxiliary being in state $\ket{g}$ with $k\in\left\{g,e,f\right\}$ as a function of the coherent $\ket{f0_\mathrm{r}}\leftrightarrow\ket{g1_\mathrm{r}}$ (d) drive frequency and (e) drive duration $T=\Theta+\tau$ for $\tau=20$\,ns when preparing the auxiliary in the second excited state $\ket{f}$.
                In~(d), the plain line shows a Lorentzian fit to the spectrum, from which the optimal drive frequency is obtained.
                In~(e), the plain line shows the fit of Eq.\,\eqref{eq:f0g1_swap} to the data from which the optimal drive duration is obtained.
                }
                \label{fig:auxiliary_reset_calibration}
            \end{figure*}

            The excited-state promoted single-shot readout previously presented is not well suited in calibration and benchmarking experiments which require the distinction between the first and second excited states of the auxiliary. In averaged measurements, it was shown in the previous subsection that the different auxiliary states can be well resolved without the excited-state promoted readout technique. When calibrating the auxiliary reset protocol with the protocol shown in Fig.\,\ref{fig:auxiliary_reset_calibration}(a), it is however useful to further convert the readout signal to approximate probabilities to be in the first excited state $\ket{e}$ and second excited state $\ket{f}$, $\tilde p_e$ and $\tilde p_f$ respectively.

            The method used to obtain these approximate probabilities consists in performing an area-preserving transforming $V\rightarrow V'$ on the raw data $V$ such that $V_g\rightarrow0$, $V_e\rightarrow V_\mathrm{eff}$, and $V_f\rightarrow iV_\mathrm{eff}$, where $V_\mathrm{eff}^2/2$ is defined as the area in the complex plane of the triangle made by the raw demodulated signals $V_g$, $V_e$, and $V_f$. Figures\,\ref{fig:auxiliary_reset_calibration}(b) and (c) show an example of the transformation.

            After transformation, the approximate probabilities corresponding to the transformed averaged readout signal $V'$ are simply estimated from $\tilde p_e=\mathrm{Re}\left(V'\right)/V_\mathrm{eff}$ and $\tilde p_f=\mathrm{Im}\left(V'\right)/V_\mathrm{eff}$, and $\tilde p_g=1-\tilde p_e-\tilde p_f$. Examples of measurements using this conversion to approximate probabilities are shown in Figure~\ref{fig:auxiliary_reset_calibration}(d--e). This method effectively normalizes out state preparation and measurement errors and can therefore not be used when characterizing initialization errors (equilibrium excited state population, Sec.\,\ref{sec:thermal_population}) or readout errors (Sec.\,\ref{ssec:readout_errors}).

        \subsubsection{Calibration of the $\ket{f0_\mathrm{r}}\leftrightarrow\ket{g1_\mathrm{r}}$ coherent swap}

            For a fixed drive amplitude and duration, the $\ket{f0_\mathrm{r}}\leftrightarrow\ket{g1_\mathrm{r}}$ drive angular frequency $\omega_{\mathrm{d},f0g1}$ is swept around the nominal value of $\omega_{f0g1}\approx2\omega_\mathrm{q}+K_\mathrm{q}-\omega_\mathrm{r}$, which does not includes the ac Stark shift induced by the drive. Figure~\ref{fig:auxiliary_reset_calibration}(d) shows the auxiliary state approximate probabilities $\tilde p_k$ for $k\in\left\{g,e,f\right\}$, obtained from the area-preserving transformation previously described, when preparing the auxiliary in the second excited state $\ket{f}$. The optimal drive angular frequency $\omega_{\mathrm{d},f0g1}$, for which $p_f$ is minimal and $p_g$ is maximal, is obtained from a Lorentzian fit to the data.
            
            Given the drive amplitude and frequency, the duration of the drive $T=\Theta_{f0g1}+\tau_{f0g1}$ is changed through the plateau $\tau_{f0g1}$ for a fixed width $\Theta_{f0g1}=20$\,ns [Fig.\,\ref{fig:auxiliary_reset_calibration}(e)]. The probability $p_f$ of the auxiliary being in state $\ket{f}$ as a function of the drive duration $T_{f0g1}=\Theta_{f0g1}+\tau_{f0g1}$ is fitted to
            \begin{align}
                p_f(T)=e^{-\kappa_{1\mathrm{r}}T/2}\left|\cos{\left(\tilde g_{\kappa_{1\mathrm{r}}}T\right)}+\frac{\kappa_{1\mathrm{r}}}{4\tilde g_{\kappa_{1\mathrm{r}}}}\sin{\left(\tilde g_{\kappa_{1\mathrm{r}}}T\right)}\right|^2,
                \label{eq:f0g1_swap}
            \end{align}
            where $\tilde g_{\kappa_{1\mathrm{r}}}=\sqrt{\tilde g^2-\left(\kappa_{1\mathrm{r}}/4\right)^2}$ is the $\ket{f0_\mathrm{r}}\leftrightarrow\ket{g1_\mathrm{r}}$ coherent swap rate $\tilde g$ renormalized by the presence of resonator dissipation at rate $\kappa_{1\mathrm{r}}$\,\cite{Magnard2018}. Note that the auxiliary decay rate $\kappa_{1\mathrm{q}}\ll\kappa_{1\mathrm{r}}$ is neglected in Eq.\,\eqref{eq:f0g1_swap}.
            
            In the fit to the data of Fig.\,\ref{fig:auxiliary_reset_calibration}(e), the value of $\kappa_{1\mathrm{r}}$ is fixed to the value of Tab.\,\ref{tab:Lindbladian_parameters} determined from spectroscopy of the resonator mode assuming $\kappa_{\phi\mathrm{r}}\ll\kappa_{1\mathrm{r}}$. The value of $\tilde g/2\pi=1.641(7)$\,MHz is obtained from fitting the data of $\tilde p_f(T)$ shown in Fig.\,\ref{fig:auxiliary_reset_calibration}(e).

        \subsubsection{Reset protocol}

            The reset protocol presented in the main text does not have the $\ket{f0_\mathrm{r}}\leftrightarrow\ket{g1_\mathrm{r}}$ and $\ket{e}\leftrightarrow\ket{f}$ drives at the same time\,\cite{Egger2018}. This is in contrast to the reset protocol used in Refs.\,\cite{Magnard2018,Sunada2022}, which requires a re-calibration of the $\pi_{ef}$ pulse in the presence of the $\ket{f0_\mathrm{r}}\leftrightarrow\ket{g1_\mathrm{r}}$ drive. After the calibration of the $\ket{f0_\mathrm{r}}\leftrightarrow\ket{g1_\mathrm{r}}$ coherent swap and the $\pi_{ef}$ pulse, the only remaining parameter to optimize in the reset delay $\delta t_\mathrm{ds}$ such that $\ket{g1_\mathrm{r}}\rightarrow\ket{g0_\mathrm{r}}$.
            Here $\delta t_\mathrm{ds}=200$\,ns is chosen, corresponding to slightly more than two resonator lifetime $T_{1\mathrm{r}}=1/\kappa_{1\mathrm{r}}=92$\,ns.

        \subsubsection{Reset errors}

            As defined in the main text, the average reset error is defined as the probability to not be in the ground state $\ket{g}$, averaged over the preparation of the auxiliary states $\ket{k}$ with $k\in\left\{g,e,f\right\}$. The conversion to approximate probabilities $\tilde p_{e}$ and $\tilde p_{f}$ based on the area-preserving transformation discussed in Sec.\,\ref{ssec:probabilities_through_transformation} is used to benchmark the auxiliary reset protocol. The reset error $\varepsilon_{\mathrm{rt}|k}$ when preparing state $\ket{k}$ is then given by
            \begin{align}
                \varepsilon_{\mathrm{rt}|k}=\tilde p_{e|k}+\tilde p_{f|k}=1-\tilde p_{g|k}.
            \end{align}
            
        \subsubsection{Average reset error in the presence of a GKP logical state}

            Through the dispersion of the auxiliary levels, the always-on cross-Kerr interaction between the storage mode and the auxiliary can increase the reset errors when the storage mode is not in a Fock state. Indeed, for the storage mode in Fock state $\ket{n_\mathrm{s}}$, a simple re-calibration of the $\ket{e}\leftrightarrow\ket{f}$ and $\ket{f0_\mathrm{r}}\leftrightarrow\ket{g1_\mathrm{r}}$ drive frequencies is, in principle, sufficient to recover a reset as good as when the storage mode is in the vacuum state $\ket{0_\mathrm{s}}$. The dispersion effect is kept under control with $\left|\chi_\mathrm{sq}\right|T_{f0g1}=1.2\times10^{-2}\ll1$. Indeed the important metric is the duration of the $\ket{f0_\mathrm{r}}\leftrightarrow\ket{g1_\mathrm{r}}$ pulse, as opposed to the complete duration of the reset that includes the reset delay.

\begin{figure*}[t]
                \centering
                \includegraphics[scale=0.81]{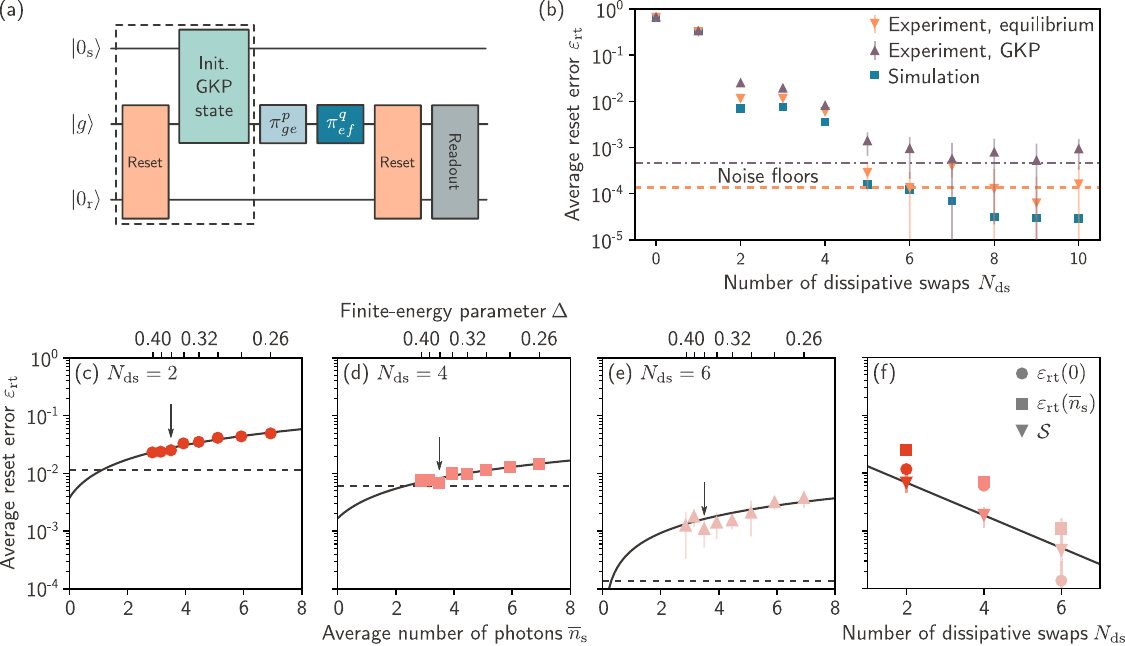}
                \caption{
                \textbf{Impact of the presence of a GKP state in the storage mode on the reset of the auxiliary.}
                (a)~Protocol for the measurement of the reset error in the presence of a GKP logical state in the storage mode. Following GKP state preparation (dashed box), the auxiliary is prepared in one of its first three levels with $\pi_{ge}^p$ and $\pi_{ef}^q$ pulses. Prior to auxiliary readout, the reset protocol of Fig.\,2(b) with $N_\mathrm{ds}$ dissipative swaps is applied. For the measurements with the storage at equilibrium\,[Fig.\,2(c)], the operations within the dashed box are omitted.
                (b)~Average reset error $\varepsilon_{\mathrm{rt}}$ as a function of the number of dissipative swaps $N_\mathrm{ds}$ when having the storage mode either at equilibrium (orange, same data as in Fig.\,2(c) of the main text) or with the GKP logical state $\ket{-\bar X}$ (purple). The simulation with the storage mode at equilibrium of Fig.\,2(c) of the main text is also shown for reference (blue). Orange dashed and purple dot-dashed lines indicate the noise floor of the measurements at equilibrium and with the GKP state, respectively.
                Average reset error $\varepsilon_{\mathrm{rt}}(\bar n_\mathrm{s})$ as a function of the average number of photons $\bar n_\mathrm{s}$ in the GKP logical state $\ket{-\bar X}$ prepared in the storage mode prior to the reset with (c)~$N_\mathrm{ds}=2$, (d)~$4$, and (e)~$6$. The average number of photons is varied by changing the finite-energy parameter $\Delta$ of the prepared GKP logical state. Plain lines show fit of Eq.\,\eqref{eq:reset_error_scaling} to the data. Dashed lines indicate average reset error $\varepsilon_{\mathrm{rt}}(0)$ measured with the storage in equilibrium [Fig.\,2(c)]. Arrows highlight results for $\Delta=0.36$, the value used throughout the manuscript.
                (f)~Average reset errors $\varepsilon_{\mathrm{rt}}(0)$ (circles) and $\varepsilon_{\mathrm{rt}}(\bar n_\mathrm{s})$ (squares) for $N_\mathrm{ds}=2$, $4$, and $6$. The scaling of the average reset error $\mathcal{S}$ from Eq.\,\eqref{eq:reset_error_scaling} is also shown (triangles). The plain line is a fit of Eq.\,\eqref{eq:reset_exponential_scaling} to the data of $\mathcal{S}$, indicating an exponential suppression of the scaling of the average reset error with the number of dissipative swaps.
                }
                \label{fig:auxiliary_reset_GKP}
            \end{figure*}

            The average reset error $\varepsilon_\mathrm{rt}$ is again measured as a function of the number of dissipative swaps $N_\mathrm{ds}$, but now with the storage mode initially prepared in the GKP $\ket{-\bar X}$ logical state with $\Delta=0.36$ with the protocol shown in Fig.\,\ref{fig:auxiliary_reset_GKP}(a). Results, shown in Fig.\,\ref{fig:auxiliary_reset_GKP}(b), indeed indicate an increase of the average reset error.

            The average reset error is also measured as a function of the average number of photons $\bar{n}_\mathrm{s}$ in the GKP $\ket{-\bar X}$ logical state through a change of the target finite-energy parameter $\Delta$ in state preparation. Results for $N_\mathrm{ds}\in\left\{2,4,6\right\}$ are shown in Fig.\,\ref{fig:auxiliary_reset_GKP}(c-e). Within a simple phenomenological model, the average reset error increases according to
            \begin{align}
                \varepsilon_{\mathrm{rt}}(\bar n_\mathrm{s})=\varepsilon_{\mathrm{rt}}(0)+\mathcal{S}\bar n_\mathrm{s},
                \label{eq:reset_error_scaling}
            \end{align}
            where $\mathcal{S}$ is a linear scaling coefficient and $\varepsilon_{\mathrm{rt}}(0)$ is the reset error with the storage in the vacuum. Is it worth noting that the initialization parameters for $\Delta=0.36$ were further optimized that those at different values of $\Delta$, most probably explaining why these results are slightly out of the general trend.

            The fitted values of $\varepsilon_{\mathrm{rt}}(0)$ appear to be lower from those measured with the storage mode at equilibrium, indicating that the simple phenomenological model is probably not valid for lower average number of photons than those measured. The scaling $\mathcal{S}$ is found to decay exponentially with the number of dissipative swaps with
            \begin{align}
                \mathcal{S}(N_\mathrm{ds})=\mathcal{S}(0)e^{-N_\mathrm{ds}/N_\mathrm{ds,0}},
                \label{eq:reset_exponential_scaling}
            \end{align}
            where $N_\mathrm{ds,0}$ is the characteristic number of dissipative swaps for the decay. The fit indicates $N_\mathrm{ds,0}=1.54(6)$. The experimental results indicate that the reset can be made exponentially robust to the presence of a GKP logical state simply by increasing the number of dissipative swaps per reset.

        \subsubsection{Numerical simulations}
        \label{ssec:reset_simulation}

            The dynamics is described by Lindblad master equation of Eq.\,\eqref{eq:displaced_Lindblad} with the driven Hamiltonian of Eq.\,\eqref{eq:driven_dispersive_Hamiltonian}. Given the $f0g1$ pulse shape, the coupling $\tilde g/2\pi=1.613$\,MHz is determined by minimizing Eq.\,\eqref{eq:f0g1_swap}. Once $\tilde g$ is determined, the maximal value of the ac Stark shift $\delta\omega_\mathrm{Stark}/2\pi=-8.94$\,MHz is calculated from Eq.\,\eqref{eq:ac_stark_shift} and the drive frequency, $\omega_{\mathrm{d},f0g1}=2\omega_\mathrm{q}+K_\mathrm{q}-\omega_\mathrm{r}+2\delta\omega_\mathrm{Stark}$, is adjusted accordingly. The amplitudes of the drives for implementing $\pi$ pulses on both $\ket{g}\leftrightarrow\ket{e}$ and $\ket{e}\leftrightarrow\ket{f}$ transitions are calibrated from analytical expressions.
            
            With all drives defined, the simulations are performed by solving the coupled differential equations resulting from Eq.\,\eqref{eq:driven_dispersive_Hamiltonian}, starting from a thermal state characterized by the equilibrium populations cited in Tab.\,\ref{tab:Lindbladian_parameters}. The average reset error is defined as in the main text and directly computed from the probability of not being in the ground state, $1-p_g$. The Hilbert space size used when the storage mode is in the vacuum state is $10\times3\times2$ for the storage, auxiliary, and resonator modes respectively.
    
\section{Gottesman-Kitaev-Preskill states}

    \subsection{Echoed conditional displacements}

\begin{figure*}[t]
            \centering
            \includegraphics[scale=0.81]{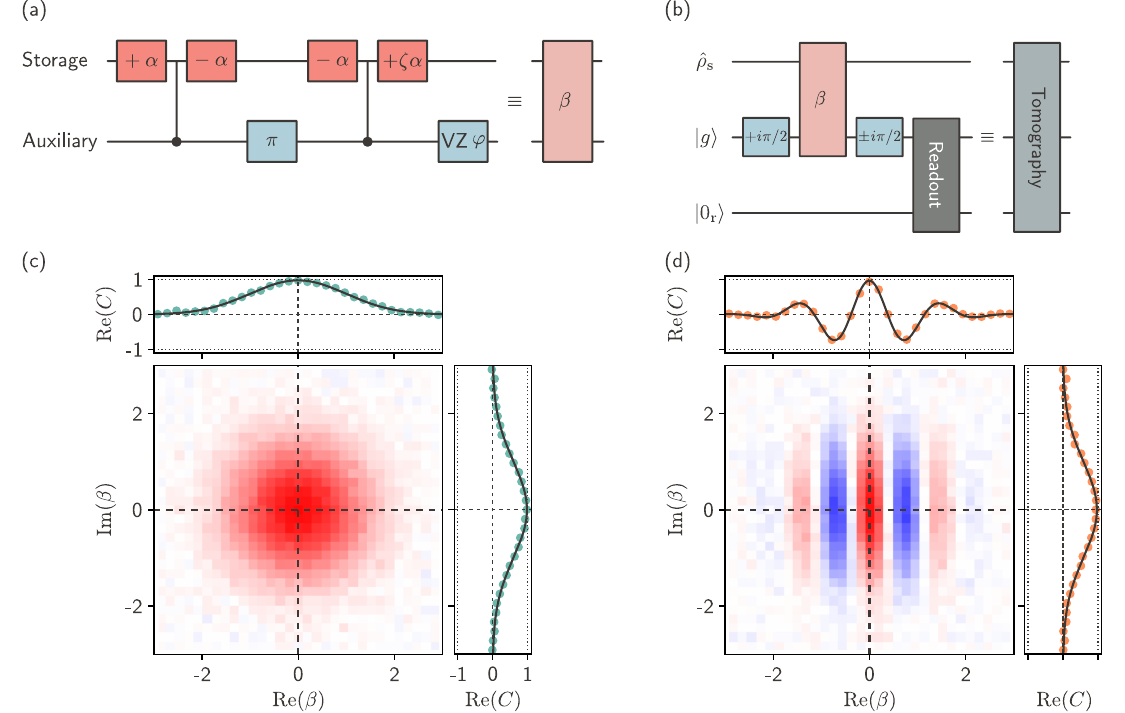}
            \caption{
            \textbf{Initial calibration of echoed conditional displacements.}
            (a)~Protocol for the implementation of echoed conditional displacements.
            (b)~Protocol to measure the real part of the characteristic function $C(\beta)$.
            Real part of the characteristic function of the storage mode prepared in (c)~the vacuum state $\ket{0_\mathrm{s}}$ and (d)~a coherent state $\ket{\lambda_\mathrm{s}}$ with $\left|\lambda\right|=2.06$. Solid line shows fit of Eq.\,\eqref{eq:vacuum_state} [\eqref{eq:coherent_state}] to the data in (c) [(d)].
            }
            \label{fig:conditional_displacement_calibration}
        \end{figure*}

        The protocol for the echoed conditional displacement (ECD), introduced in Refs.\,\cite{Campagne-Ibarcq2019a,Eickbusch2022}, is shown in Fig.\,\ref{fig:conditional_displacement_calibration}(a). The echoed conditional displacements is based on unconditional displacements of the storage mode and conditional rotation of the storage mode enabled by the storage-auxiliary cross-Kerr interaction. The last displacement pulse is scaled with parameter $\zeta$ to correct for a spurious unconditional displacement. A virtual-Z gate of parameter $\varphi$ is applied on the auxiliary at the end to correct for a finite geometric phase accumulated during the ECD.

        \subsubsection{Tomography and initial calibration}

            The protocol for the measurement of the characteristic function $C(\beta)$ is shown in Fig.\,\ref{fig:conditional_displacement_calibration}(b)\,\cite{Fluhmann2020}. The first step in the calibration procedure is to determine the approximate size of the conditional displacement $\beta$ resulting from the echoed conditional displacement pulse sequence for a given displacement pulse amplitude $A_\mathrm{s}$. This is done through the measurement of the characteristic function $C(\beta)$ of the vacuum state, given by
            \begin{align}
                C_{\ket{0_\mathrm{s}}}(\beta)=e^{-\left|\beta\right|^2/2}.
                \label{eq:vacuum_state}
            \end{align}
            The proportionality constant $c_\beta$ between $\beta$ and $A_\mathrm{s}$ is defined such that $\left|\beta\right|=c_\beta\left|A_\mathrm{s}\right|$.
            
            The calibration of the conditional displacement of size $\left|\alpha\right|$ used to generate the conditional displacement of size $\left|\beta\right|$ is done by measuring the coherent state generated by such a displacement pulse. For a coherent state $\ket{\lambda_\mathrm{s}}$ of size $\left|\lambda\right|$ generated from a pulse of amplitude $A_{\mathrm{s}\lambda}$, the characteristic function is given by
            \begin{align}
                C_{\ket{\lambda_\mathrm{s}}}(\beta)=e^{-\left|\beta\right|^2/2}\cos\left(2\left[\mathrm{Re}(\lambda)\mathrm{Im}(\beta)+\mathrm{Im}(\lambda)\mathrm{Re}(\beta)\right]\right).
                \label{eq:coherent_state}
            \end{align}
            The characteristic function of a coherent state is complex-valued, but not measuring the imaginary part only leads to an ambiguity on the sign of the coherent state, which is irrelevant for the calibration presented here. Given the previous calibration of $\left|\beta\right|$, the proportionality constant $c_\alpha$ relates the unconditional displacement of size $\left|\alpha\right|$ and the displacement pulse amplitude $A_\mathrm{s}$ with $\left|\alpha\right|=c_\alpha\left|A_\mathrm{s}\right|$ where $c_\alpha=\left|\lambda\right|/\left|A_{\mathrm{s}\lambda}\right|$. Figures\,\ref{fig:conditional_displacement_calibration}(c-d) show examples of the measurement of the real part of $C(\beta)$ for vacuum and coherent states.

            The estimation of the proportionality constant $c_\beta$ presented here is further fine-tuned in closed-loop optimizations presented in Sec.\,\ref{ssec:closed-loop_ECD}. Furthermore, based on the numerical calculation presented in Sec.\,\ref{sssec:ECD_simulations}, the nonlinearity originating from the storage mode self-Kerr of coefficient $K_\mathrm{s}$ and storage-auxiliary second-order cross-Kerr of coefficient $\chi_\mathrm{s^2q}$ is accounted for when calculating the required pulse amplitude $A_\mathrm{s}$ to achieve a target $\beta$.

            For the tomography of the GKP logical states, only the real part of the characteristic function is measured and the imaginary part is assumed to be zero, \textit{\textit{i.e.}} $C(\beta)=\mathrm{Re}(C(\beta))$. As later discussed, the spurious geometric phase accumulated by the auxiliary during tomography is corrected to prevent the real and imaginary parts of the characteristic function to be mixed. This is in contrast of Ref.\,\cite{Eickbusch2022} in which the spurious auxiliary phase is not corrected during tomography; instead the imaginary part is measured in order to reconstruct $C(\beta)$. However, in Ref.\,\cite{Eickbusch2022}, the symmetry $C(\beta)=C(-\beta)$ is enforced by measuring only for $\mathrm{Re}(\beta)\geq0$, such that the number of measurement is equivalent in both approaches.
   
        \subsubsection{Calibration of storage drive frequency and storage-auxiliary cross-Kerr}
        \label{ssec:initial_calibration}

            The calibration of the storage drive frequency $\omega_\mathrm{d,s}$ for echoed conditional displacements and the storage-auxiliary cross-Kerr coefficient $\chi_\mathrm{sq}$ are obtained by tracking the phase of a small coherent state after a finite free evolution time. The state angle is obtained by measuring the characteristic function $C_{\ket{\lambda}}(\beta)$ of a coherent state $\ket{\lambda}$ evolving for a fixed time $T=1.774\,\unit{\micro\second}$ as a function of the storage drive frequency.

\begin{figure*}[t]
                \centering
                \includegraphics[scale=0.81]{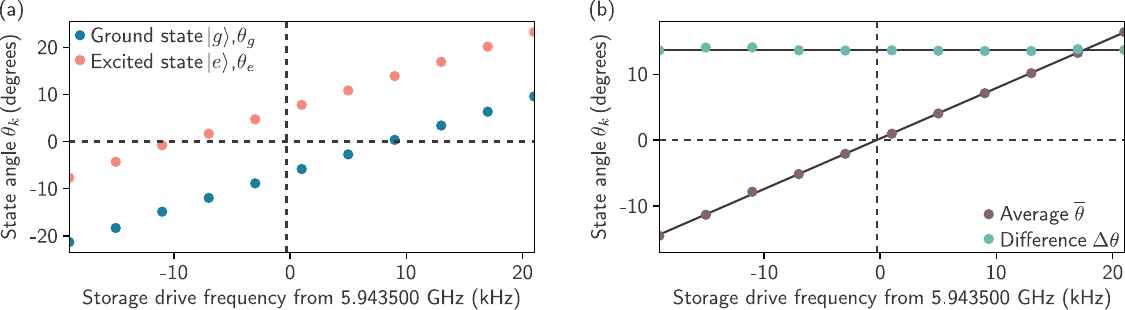}
                \caption{
                \textbf{Calibration of the storage drive frequency and storage-auxiliary cross-Kerr coefficient.}
                (a)~State angle $\theta_k$ of a coherent state $\ket{\lambda_\mathrm{s}}$ with $\lambda=\left|\lambda\right|e^{i\theta_k}$ and $\left|\lambda\right|=2.98$ after a free evolution time of $T=1.774$\,\unit{\micro\second} when initially preparing the auxiliary in state $\ket{k}$ with $k\in\left\{g,e\right\}$ as a function of the storage drive frequency.
                (b)~Average state angle $\overline\theta=(\theta_g+\theta_e)/2$ and difference $\Delta\theta=\theta_e-\theta_g$ of state angles as a function of the storage drive frequency. The vertical dashed lines indicate the drive frequency for which $\overline\theta=0$ (horizontal dashed lines). In (b), the horizontal plain line indicates the average value of the average state angle $\bar\theta$, and the plain line is a linear fit to the data of difference $\Delta\theta$ of state angles.
                }
                \label{fig:displacement_frequency_calibration}
            \end{figure*}

            Figure~\ref{fig:displacement_frequency_calibration}(a) shows the state angle with the free evolution performed with the auxiliary prepared in either the ground state $\ket{g}$ or excited state $\ket{e}$. Figure~\ref{fig:displacement_frequency_calibration}(b) shows the difference in the state angle $\Delta\theta=\theta_e-\theta_g$, proportional to the storage-auxiliary dispersive shift coefficient $\chi_\mathrm{sq}$ with $\Delta\theta=-2\chi_\mathrm{sq}T$. Furthermore, the drive frequency desired for echoed conditional displacements is found as the one such that the state rotation averaged over the auxiliary states $\ket{g}$ and $\ket{e}$ cancels out.

        \subsubsection{Corrections of spurious displacements and phases}

            As shown in Fig.\,\ref{fig:conditional_displacement_calibration}(a), the last displacement pulse of the ECD protocol is scaled with a factor $\zeta$ to correct for the spurious displacement perpendicular to the conditional displacement\,\cite{Campagne-Ibarcq2019a,Eickbusch2022}. Furthermore, the spurious geometric phase accumulated by the auxiliary during an ECD is corrected through a virtual-Z gate on the auxiliary\,\cite{Eickbusch2022}. The spurious phase $\varphi$ is considered to scale quadratically with pulse amplitude $\left|A_\mathrm{s}\right|$, \textit{i.e.} $\varphi=c_\varphi\left|A_\mathrm{s}\right|^2$. The parameterization as a function of pulse amplitude as opposed to conditional displacement amplitude $\left|\beta\right|$ is to make the calibration of $c_\varphi$ independent on the calibration of $c_\beta$. Both $\zeta$ and $c_\varphi$ are obtained from closed-loop optimizations, to be later discussed, for each protocols using echoed conditional displacements (Sec.\,\ref{ssec:closed-loop_ECD}).

        \subsubsection{Measurement of the storage mode relaxation time}
        \label{ssec:storage_relaxation}

            The relaxation time of the storage mode needs to be measured in a phase-insensitive manner. By definition, the only phase-insensitive point in phase space is at $\beta=0$. However, $C(\beta=0)=1$ by definition for all states, and therefore yields no information on the state of the storage mode. Hence, there is no single-point measurement of the characteristic function that is phase-insensitive and yield information about the state. One way to get back a phase-insensitive information from the characteristic function is to integrate over phase space. Indeed, the expectation value of the photon-number-parity operator $\mathcal{\hat P}=e^{i\pi\hat a^\dagger\hat a}$ is given by,
            \begin{align}
                \langle\mathcal{\hat P}\rangle=\frac{1}{2\pi}\int\ \mathrm{d}^2\beta\ C(\beta)=\frac{1}{2\pi}\int\mathrm{d}^2\beta \ \mathrm{Re}\left(C(\beta)\right),
            \end{align}
            where the second equality comes from the fact that $\langle\mathcal{\hat P}\rangle$ is real\,\cite{Haroche2006}. In other words, the expectation value of the parity operator, or simply parity hereafter, is obtained from the integral of the characteristic function. One advantage of this method is that it is insensitive to a possible miscalibration of the conditional displacements as long as the range for integration is large enough.
 
\begin{figure*}[t]
                \centering
                \includegraphics[scale=0.81]{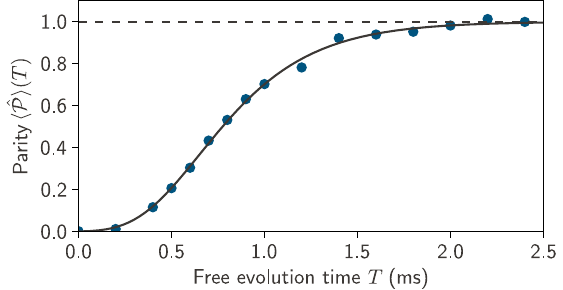}
                \caption{
                \textbf{Relaxation of the storage mode}. The expectation value of the parity operator, $\langle\mathcal{\hat P}\rangle$, is measured as a function of the free evolution time $T$. The plain line shows the fit of Eq.\,\eqref{eq:Poisson_decay} to the data, yielding $T_\mathrm{1s}=0.335$\,ms. The dashed line indicates the steady-state value of $\langle\mathcal{\hat P}\rangle=1$.
                }
                \label{fig:storage_relaxation}
            \end{figure*}

            The storage mode relaxation time $T_{1\mathrm{s}}$ is measured from the relaxation of a small coherent state $\ket{\lambda_0}$ as opposed to a single Fock state $\ket{1_\mathrm{s}}$. Figure~\ref{fig:storage_relaxation} shows the measurement $\langle\mathcal{\hat P}\rangle(T)$ for a small coherent state $\ket{\lambda_0}$ with $\left|\lambda_0\right|=1.868(11)$ as a function of the free evolution time $T$. The parity decays according to
            \begin{align}
                \langle\mathcal{\hat P}\rangle(T)=\mathrm{exp}\left(-2\left|\lambda_0\right|^2e^{-T/T_{1\mathrm{s}}}\right).
                \label{eq:Poisson_decay}
            \end{align}
            The only fitting parameter is the storage mode relaxation time, which yields $T_\mathrm{1s}=0.335\,\unit{\milli\second}$.

        \subsubsection{Numerical simulations}
        \label{sssec:ECD_simulations}

            For the numerical simulation of an echoed conditional displacement of amplitude $\beta$, the drive amplitude $\epsilon_\mathrm{s0}$ is first estimated with the expression
            \begin{align}
                \epsilon_\mathrm{s0}=\beta\chi_\mathrm{sq}/(4i\left[\cos(\chi_\mathrm{sq}(\Theta_\mathrm{s} + \tau_\mathrm{s}+\delta t_\mathrm{ECD}))-\cos(\chi_\mathrm{sq}(2(\Theta_\mathrm{s} + \tau_\mathrm{s})+\delta t_\mathrm{ECD}))+\cos(\chi_\mathrm{sq}(\Theta_\mathrm{s} + \tau_\mathrm{s}))-1\right]),
                \label{eq:ECD_amplitude_estimate}
            \end{align}
            where $\Theta_\mathrm{s}$ and $\tau_\mathrm{s}$ are respectively the storage drive width and plateau, and $\delta t_\mathrm{ECD}$ is the spacing between the first and second displacement pulses, and third and fourth displacement pulses [Fig.\,\ref{fig:conditional_displacement_calibration}(a)].
            
            Corrections are applied to this estimation by numerically computing the value of $\beta$ and checking the difference with the target $\beta$. This numerical calculation is based on
            \begin{align}
                \beta=2i\int_{T_\mathrm{i}}^{T_\mathrm{f}}\mathrm{d}t\ e^{-i \Delta_\mathrm{eff}(t)(t-T_\mathrm{f})}\Biggr[\chi_\mathrm{eff}(t) s(t)\cos(\chi_\mathrm{eff}(t)\int_{t}^{T_\mathrm{f}}\mathrm{d}t'\ s(t'))-i\Delta_\mathrm{eff}(t)\sin(\chi_\mathrm{eff}(t) \int_{t}^{T_\mathrm{f}}\mathrm{d}t'\ s(t'))\Biggr]\alpha(t),
            \end{align}
            where the effective detuning is given by
            \begin{align}
                \Delta_\mathrm{eff}(t)=\Delta_\mathrm{s}-\chi_\mathrm{sq}-2K_\mathrm{s}\abs{\alpha(t)}^2,
            \end{align}
            and the effective cross-Kerr by
            \begin{align}
                \chi_\mathrm{eff}(t)=\chi_\mathrm{sq}+2\chi_\mathrm{s^2q}\abs{\alpha(t)}^2,
                \end{align}
            where $\alpha(t)=-i\int_{T_\mathrm{i}}^{t}\mathrm{d}t'\ \epsilon_\mathrm{s}(t')$ is the storage unconditional displacement and $s(t)=\mathrm{sign}(T_\pi-t)$ models the auxiliary $\pi$-pulse at $t=T_\pi$. Effective detuning and cross-Kerr, $\Delta_\mathrm{eff}(t)$ and $\chi_\mathrm{eff}(t)$, are introduced to account for nonlinear effects proportional to $\abs{\alpha(t)}^2$ coming from the storage self-Kerr $K_\mathrm{s}$ and the second-order storage-auxiliary cross-Kerr $\chi_\mathrm{s^2q}$. This correction process is repeated until the difference becomes lower than a given threshold. This method allows numerical simulations of echoed conditional displacements with $\sim10^{-4}$ infidelity.

    \subsection{Initialization}

        \subsubsection{Initialization parameters}

            Candidates of initialization parameters $\left\{\mathbf{R},\bm{\beta}\right\}$ for GKP logical states are obtained with gate-level simulations with \textit{tensorflow}. Each gate corresponds to a layer in a sequential neural network. The optimization is performed using gradient descent and the cost function is taken as $(1-F_{\hat\rho})^2$, where $F_{\hat\rho}$ is the state fidelity as defined in the main text. For a given finite-energy parameter $\Delta$ and GKP cardinal state $\ket{-\bar X}$, $\ket{+\bar X}$, or $\ket{-\bar Y}$, we optimize batches of $100$ circuits for $250$ epochs, and repeat this $50$ times, totalling $5000$ optimized circuits per state. The best gate-level candidates are then tested in an open-system, pulse-level numerical simulations. Circuit of depth $N=9$ is found to be close to optimal in simulations given the noise of the system.
            
            As previously discussed, the measurement of the characteristic function of the cardinal states $\ket{-\bar X}$, $\ket{+\bar X}$, and $\ket{-\bar Y}$ is sufficient to benchmark the initialization of the GKP qubit given that the states $\ket{-\bar Z}$, $\ket{+\bar Z}$ and $\ket{+\bar Y}$ are respectively identical up to a $90^\circ$ rotation. The information about these states is therefore all included in the measurement of the characteristic function as presented in Fig.\,1(d--f) of the main text.

        \subsubsection{Closed-loop optimization of ECD parameters}
        \label{ssec:closed-loop_ECD}
           
            The optimization parameters related to echoed conditional displacements used in the initialization and tomography protocols are the control parameters $c_\beta$, $\zeta$, and $c_\varphi$ previously introduced. The nonlinearity correction is fixed through the Hamiltonian parameters and only the proportionality constant $c_\beta$ is used as an optimization parameter. For the initialization protocol, a uniform rotation of angle $\Delta\theta_\mathrm{init}$ with $\beta_n\rightarrow\beta_n e^{i\Delta\theta_\mathrm{init}}$ on the values of $\beta_n$ is also used as an optimization parameter to compensate for the finite state rotation during the auxiliary reset. The optimization of the ECD parameters for the quantum error correction protocol is discussed in Sec.\,\ref{ssec:closed-loop_QEC}.
            
            For the tomography protocol, the ECD control parameters need to be the same for all cardinal states. Here, the choice is made to look at a particular cardinal state, $\ket{-\bar X}$. For initialization, the optimal parameters can be different for every cardinal states. After the state rotation correction $\Delta\theta_\mathrm{init}$, for which the optimal value is state-dependent, it is important that the initial angle is the same for all states given that the quantum error correction parameters need to be state-independent.
    
\begin{figure*}[t]
                \centering
                \includegraphics[scale=0.81]{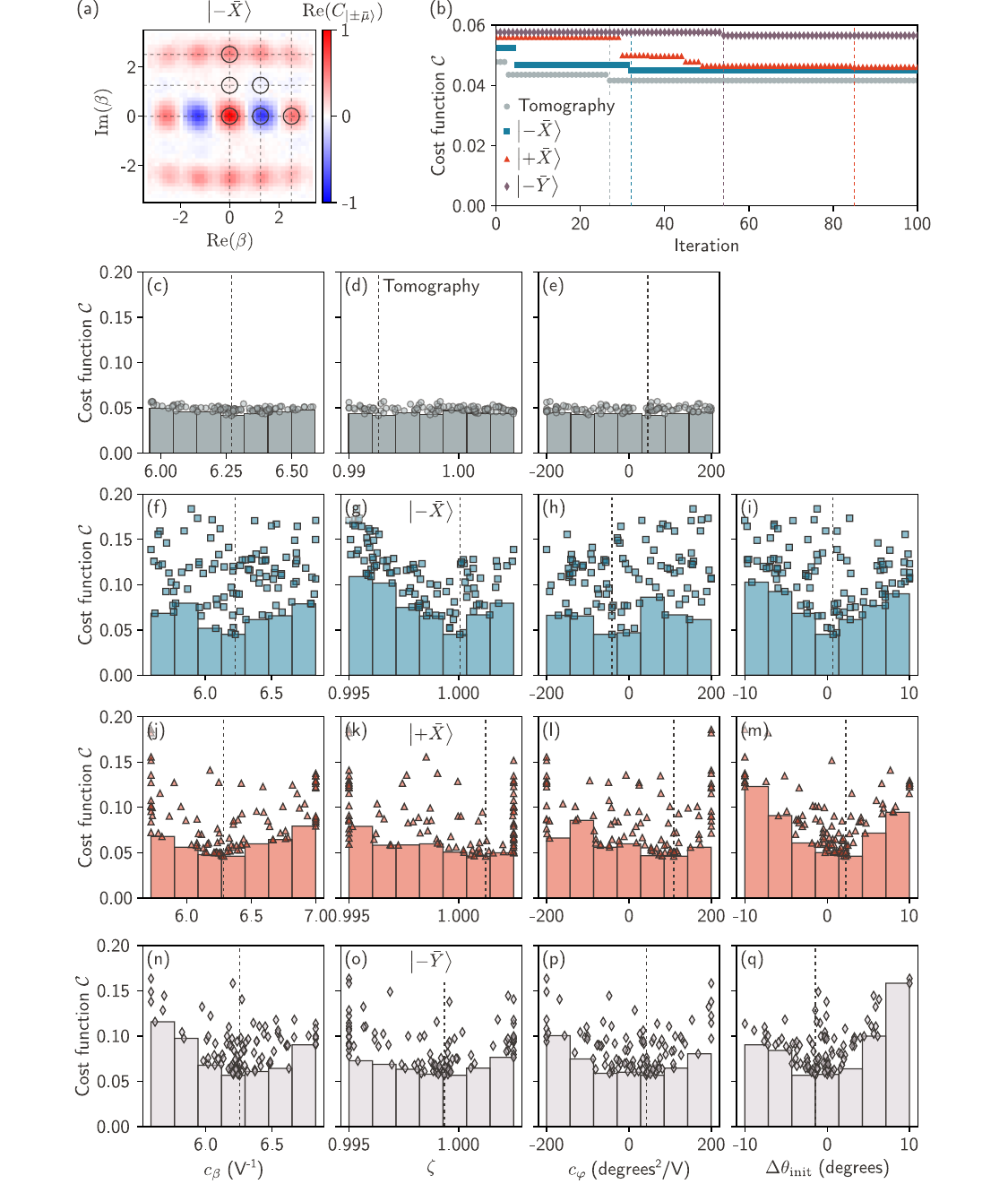}
                \caption{
                \textbf{Closed-loop optimization of ECD parameters.}
                (a)~Real part of the characteristic function, $\mathrm{Re}(C_{\ket{\pm\bar\mu}})$, of GKP logical state $\ket{-\bar X}$ (same data as in Fig.\,1(d) of the main text). The black circles indicates the values of $\beta_m$ at which the characteristic function is measured to evaluate the cost function $\mathcal{C}$ defined in Eq.\,\eqref{eq:cost_function_ECD}. The vertical and horizontal dashed lines respectively indicate $\mathrm{Re}(\beta)$ and $\mathrm{Im}(\beta)$ at $\left\{0,1/2,1\right\}\times\ell/\sqrt{2}$.
                (b)~Cumulative minimal value of the cost function $\mathcal{C}$ for the tomography and initialization protocols.
                Cost function $\mathcal{C}$ as a function of the (c,f,j,n) conditional displacement proportionality constant $c_\beta$, (d,g,k,o) spurious displacement correction factor $\zeta$, and (e,h,l,p) spurious auxiliary phase scaling $c_\varphi$ for the (c--e) tomography protocol, (f--i) initialization of $\ket{-\bar X}$, (j--m) $\ket{+\bar X}$, and (n--q) $\ket{-\bar Y}$.
                The cost function $\mathcal{C}$ as a function of the state rotation correction $\Delta\theta_\mathrm{init}$ is also shown for initialization of (i) $\ket{-\bar X}$, (m) $\ket{+\bar X}$, and (q) $\ket{-\bar Y}$.
                Data points show the results from all iterations, while bars indicate the minimal value of the cost function within each bins.
                Dashed lines indicate the optimal value for each parameter for each protocol.
                }
                \label{fig:closed-loop_optimization_ECD}
            \end{figure*}

            To perform the closed-loop optimization, the real part of the characteristic function is measured at the six values of $\beta$ shown in Fig.\,\ref{fig:closed-loop_optimization_ECD}(a). Measurements of the Pauli expectation values $\expval{\hat\mu_0}$ and measurements of the expectation values for the two infinite-energy stabilizers $\hat S_0^X$ and $\hat S_0^Z$ inform on the state of the GKP qubit. In addition to these five measurements, a measurement at $\beta=0$ informs on the purity of the auxiliary. More explicitly, the characteristic function is measured at $\beta_m\in\left\{0,\beta_{\hat X_0},\beta_{\hat Y_0},\beta_{\hat Z_0},\beta_{\hat S_0^X},\beta_{\hat S_0^z}\right\}$, with $\beta_{\hat X_0}=\ell/\sqrt{8}$, $\beta_{\hat Y_0}=(1+i)\ell/\sqrt{8}$, $\beta_{\hat Z_0}=i\ell/\sqrt{8}$, $\beta_{\hat S_0^X}=\ell/\sqrt{2}$, and $\beta_{\hat S_0^Z}=i\ell/\sqrt{2}$.

            Based on these measurements, the cost function is defined as the averaged square root of the squared difference between the experiment result and the expected value for an ideal finite-energy state. The average is over the six different measurement settings. More explicitly,
            \begin{align}
                \mathcal{C}=\frac{1}{6}\sqrt{\sum_{\beta_m}\left(\mathrm{Re}(C_{\ket{\pm\bar\mu}}(\beta_m))-\mathrm{Re}(C_{\ket{\pm\bar\mu}}^\mathrm{target}(\beta_m))\right)^2},
                \label{eq:cost_function_ECD}
            \end{align}
            where $\mathrm{Re}(C_{\ket{\pm\bar\mu}}(\beta_m))$ [$\mathrm{Re}(C_{\ket{\pm\bar\mu}}^\mathrm{target}(\beta_m))$] is the experimentally measured [ideal] value of the real part of the characteristic function measured at $\beta_m$ for $\beta_m\in\left\{0,\beta_{\hat X_0},\beta_{\hat Y_0},\beta_{\hat Z_0},\beta_{\hat S_0^X},\beta_{\hat S_0^z}\right\}$.
            The minimal value of the cost function is $\mathcal{C}=0$ for an ideal state in the absence of readout errors. To approximately factor out auxiliary state preparation and readout errors, each value of $\mathrm{Re}(C_{\ket{\pm\bar\mu}}^\mathrm{target}(\beta_m))$ are scaled with the readout visibility $\mathcal{V}$.

            The optimizer used is \texttt{GaussianProcess} from Boulder Opal by Q-CTRL. The number of iterations is $100$, plus an initial guess from a previous optimization. Results for the tomography and initialization protocols are shown in Fig.\,\ref{fig:closed-loop_optimization_ECD}(b--q). The optimal values of the proportionality constant $c_\beta$ differ by $0.8\%$ across the five optimizations ($95\%$ confidence interval), including the optimization for the QEC protocol discussed in Sec.\,\ref{ssec:closed-loop_QEC}. Optimal values of the scaling factor $\zeta$ both smaller and greater than $1$ are obtained, with an average value $\zeta-1=-1.1(6.4)\times10^{-3}$. To put this number in perspective, given $\left|\beta\right|/\left|\alpha\right|=0.0804$, the \textbf{Big} conditional displacement of amplitude $|\beta_\mathbf{B}|=\ell/\sqrt{2}\approx2.51$ requires an unconditional displacement $|\alpha_\mathbf{B}|\approx31.2$ during the ECD. Therefore, the averaged $\zeta-1$ leads to an unconditional displacement correction of amplitude $\Delta|\alpha_\mathbf{B}|\approx-0.034(199)$.

        \subsubsection{Numerical simulations}
        \label{sssec:init_simu}

            The set of echoed conditional displacements in the initialization protocol is constructed following the method in Sec.\,\ref{sssec:ECD_simulations}, where the control parameters $\zeta$ and $c_\varphi$ are left free for optimization. Those two parameters are optimized using the \texttt{scikit-optimize} package with the following cost function
            \begin{align}
                \mathcal{C}=\sum_{\hat\mu_0} (\expval{\hat\mu_0}-\expval{\hat\mu_0}_\mathrm{target})^2,
            \end{align}
            where $\expval{\hat\mu_0}$ ($\expval{\hat\mu_0}_\mathrm{target}$) is the expectation value of the infinite-energy Pauli operator or stabilizer $\hat\mu_0\in\left\{\hat X_0,\hat Y_0,\hat Z_0,\hat S_0^X,\hat S_0^Z\right\}$ for a given initialized (ideal) state. This closed-loop optimization process takes initialized states from simulations using Hilbert space dimensions of $60$, $2$ and $1$ for the storage, auxiliary and resonator modes respectively, and the reset was not simulated at the end of these numerical protocols.
        
            Using the optimal parameters obtained from this method, simulations of the initialization protocol, including the reset, are performed with Hilbert space sizes of $60$, $3$ and $2$ for the storage, auxiliary and resonator modes respectively, for preparing the three GKP logical states $\ket{-\bar X}$, $\ket{+\bar X}$, and $\ket{-\bar Y}$ with $\Delta=0.36$. These simulations are performed as a function of the intrinsic dephasing rate $\kappa_{\mathrm{s}\phi}$ of the storage mode. The value of $\kappa_{\mathrm{s}\phi}=(110\,\unit{\milli\second})^{-1}$, consistent with Ref.\,\cite{Eickbusch2022}, is estimated by minimizing the sum of the squared difference between the state fidelity from experiments and simulations for the three GKP states (Tab.\,\ref{tab:Lindbladian_parameters}).

            The density matrices of these initialized states are taken as input states for the numerical simulation of the tomography protocol. The real part of the characteristic function $\mathrm{Re}(C_{\ket{\pm\bar\mu}}(\beta))$ is evaluated from the expectation value of $\hat\sigma_z$ of the auxiliary. However, as previously discussed, the auxiliary readout is not simulated. Therefore, to account for readout errors, the expectation values $\expval{\hat\sigma_z}$ from numerical simulations are rescaled considering the probability $\varepsilon_g$ of mistakenly getting the outcome $-1$ given the state is $\ket{g}$, and vice versa for $\varepsilon_e$ (Sec.\,\ref{ssec:readout_errors}).

            With this simulated measurement of the characteristic function, the density matrices are reconstructed using the same method used for experimental results. The numerical results, based on reconstructed density matrices, are used in the comparison with the experimental results discussed in the main text. The density matrices obtained directly from the state preparation protocol, \textit{i.e.} without reconstruction, are used as the input for simulations used to obtain the logical lifetime in simulation.

        \subsubsection{Error budget}
        \label{sssec:init_error_budget}
        
        Table~\ref{tab:error_budget} shows the error budget for the initialization error $1-F_{\hat\rho}$ averaged over the cardinal states. The error budget is decomposed into error channels from both the storage and auxiliary modes,  and from coherent control. Values are obtained from numerical simulations of the initialization protocol without one of the error channel and comparing the average infidelity with the results from simulations with all error channels.

        \begin{table*}[t]
        \centering
        \begin{tabular}{l||l||c}
            Physical mode           & Error channel         & State error           \\
                                    &                       & $1-F_{\hat\rho}$ (\%) \\
            \hline\hline
            Storage mode            & Single-photon loss    & $2.02$                \\
                                    & Dephasing             & $6.40$                \\
                                    & Self-Kerr             & $0.03$                \\
            \hline
            Auxiliary               & Single-photon loss    & $4.00$                \\
                                    & Dephasing             & $0.62$                \\
                                    & Equilibrium population& $0.07$                \\
            \hline
                                    & Coherent control      & $0.37$                \\
            \hline\hline
                                    & Sum                   & $13.52$               \\
                                    & All                   & $14.73$               \\
        \hline\hline
                                    & Experiment            & $14.69$               \\
        \end{tabular}
        \caption{
        \textbf{Error budget} for the state error $1-F_{\hat\rho}$ after initialization averaged over the GKP qubit cardinal states from different error channels.
            }
        \label{tab:error_budget}
        \end{table*}     
            
        Error channels are removed by setting to zero their corresponding parameter in the Lindblad master equation. From this method, the effect of equilibrium populations are also contained in the single-photon loss error channel. Coherent control errors are evaluated differently by simply removing all the other error channels and taking the resulting average infidelity. For the purpose of the error budget, the infidelity is computed directly on the states obtained from numerical simulations of the initialization protocol.
        
        Table~\ref{tab:error_budget} also includes the sum of the individual error channels, as well as the average infidelity from full simulations with all error channels. The difference between the sum and the full simulations can come from cascaded errors from the combination of multiple error channels. Table~\ref{tab:error_budget} shows that initialization errors are dominated by auxiliary relaxation and storage mode dephasing amplified during conditional displacements.

    \subsection{Logical fidelity and lifetime}

        As in the main text, the logical fidelity is given by 
            \begin{align}
                F_\mathrm{L}=\frac{1}{2}+\frac{1}{12}\sum_{\hat\mu_0\in\{\hat X_0,\hat Y_0,\hat Z_0\}}\left(\expval{\hat\mu_0}_+-\expval{\hat\mu_0}_-\right),
               \label{eq:logical_channel_fidelity}
            \end{align}
        where $\expval{\hat\mu_0}_\pm=\expval{\pm\bar\mu|\hat\mu_0|\pm\bar\mu}$ is the expectation value of the infinite-energy Pauli operator $\hat\mu_0$ when a logical state $\ket{\pm\bar\mu}$ with $\bar\mu\in\left\{\bar X,\bar Y,\bar Z\right\}$ is prepared\,\cite{Nielsen2002,Sivak2023}. The Pauli expectation value corresponds to the real part of the characteristic function $C_{\ket{\pm\bar\mu}}(\beta_{\hat\mu_0})$ of the state $\ket{\pm\bar\mu}$ with $\beta_{\hat X_0}=\ell/\sqrt{8}$, $\beta_{\hat Y_0}=(1+i)\ell/\sqrt{8}$ and $\beta_{\hat Z_0}=i\ell/\sqrt{8}$ with $\ell=2\sqrt{\pi}$ for the square GKP qubit\,\cite{Terhal2020a,Royer2020}.
        
        The evaluation of the Pauli expectation value $\expval{\hat\mu_0}_\pm=\expval{\pm\bar\mu|\hat\mu_0|\pm\bar\mu}$ with $\hat\mu_0\in\{\hat X_0,\hat Y_0,\hat Z_0\}$ requires the measurement of $\langle\hat X_0\rangle_\pm$ and $\langle\hat Y_0\rangle_\pm$. Indeed, $\langle\hat X_0\rangle_\pm=\langle\hat Z_0\rangle_\pm$ given that both the conditional displacements of complex amplitude $\beta_n$ used for state preparation and the complex amplitude of the measurement of the characteristic function $\beta_{\hat X_0}$ are transformed with $\beta\rightarrow i\beta$ for $\langle\hat X_0\rangle_\pm\rightarrow\langle\hat Z_0\rangle_\pm$. Therefore, to evaluate the logical fidelity of Eq.\,\eqref{eq:logical_channel_fidelity}, measurements of only four out of six values of $\expval{\hat\mu_0}_\pm$ are required, $\langle\hat X_0\rangle_\pm$ and $\langle\hat Y_0\rangle_\pm$.

       Two different methods used to obtain the logical lifetime from either experimental or simulation data is presented in Sec.\,\ref{ssec:method_A} and \ref{ssec:method_B} and later compared.

        \subsubsection{Measurement of the Pauli expectation values and finite-energy parameter}
        \label{ssec:Pauli_measurement}

            A single-point measurement of the real part of the characteristic function, $\mathrm{Re}\left(C_{\ket{\pm\bar\mu}}(\beta_{\hat\mu_0})\right)$, can underestimate the Pauli expectation value $\expval{\hat\mu_0}_\pm=\expval{\pm\bar\mu|\hat\mu_0|\pm\bar\mu}$ if the phase of $\beta_{\hat\mu_0}$ is miscalibrated, \textit{\textit{i.e.}} if the frame tracking is not perfect. To make the measurement of the Pauli expectation value more robust, the characteristic function is measured as a function of the phase $\mathrm{arg}(\beta)$ at a fixed amplitude $\left|\beta\right|=|\beta_{\hat\mu_0}|$. The phase $\mathrm{arg}(\beta)$ is swept by $\pm30^\circ$ around the expected state angle at a given time for a given protocol.

            The characteristic function of an ideal GKP state of finite-energy parameter $\Delta$, described by the density matrix $\hat\rho_\mathrm{target}=\ket{\pm\bar\mu}\bra{\pm\bar\mu}$, is numerically evaluated with $C_{\ket{\pm\bar\mu}}(\beta_{\hat\mu_0})=\langle{\hat D(\beta_{\hat\mu_0})}\rangle=\mathrm{Tr}\left[\hat\rho_\mathrm{target}e^{\beta\hat a^\dagger-\beta^*\hat a}\right]$, where $\hat D(\beta)$ is the displacement operator\,\cite{Haroche2006}. This numerical calculation is fitted to the experimental data with the fitting parameters being the finite-energy parameter $\Delta$, the state angle $\theta$, and an overall contrast $\eta$ to account for state preparation and measurement errors. This method enables one to track the time-evolution of the finite-energy parameter $\Delta$.

            While the Pauli expectation values can be estimated from the fit directly, it is found to be unreliable when the signal-to-noise ratio is too low and $\Delta$ gets too large in the case without QEC. Therefore, the measurement of the Pauli expectation values are taken from a single point interpolation of the data at a phase $\mathrm{arg}(\beta)$ determined from the state rotation, which is obtained from a linear fit to the values of $\theta(T)$ obtained from the numerical fit previously described.

        \subsubsection{Time-dependence of Pauli expectation values}

            The Pauli expectation value for each state decays exponentially with time $T$ after initialization according to
            \begin{align}
                \expval{\hat\mu_0}_\pm(T)=\expval{\hat\mu_0}_\pm(0)e^{-T/T_{\ket{\pm\bar\mu}}}+f_{\ket{\pm\bar\mu}}(T),
                \label{eq:time_dependence_Pauli}
            \end{align}
            where $T_{\ket{\pm\bar\mu}}$ is the lifetime of logical state $\ket{\pm\bar\mu}$ and $f_{\ket{\pm\bar\mu}}(T)$ is a possible time-dependent offset with $f_{\ket{\pm\bar\mu}}(0)=0$\,\cite{Sivak2023}. For example, for the case without QEC, the time-dependent offset comes from the fact that Pauli expectation values have non-zero values at equilibrium, \textit{i.e.} when very close to the vacuum state $\ket{0_\mathrm{s}}$. Indeed, the characteristic function of the vacuum state, $C_{\ket{0_\mathrm{s}}}(\beta)=e^{-\left|\beta\right|^2/2}$ is nonzero for any finite value of $\left|\beta\right|$. More explicitly,
            \begin{align}
                \langle\hat X_0\rangle_\pm(T\rightarrow\infty)&=\mathrm{Re}\left(C_{\ket{0_\mathrm{s}}}(\beta_{\hat X_0})\right)=e^{-\pi/4},\nonumber\\
                \langle\hat Y_0\rangle_\pm(T\rightarrow\infty)&=\mathrm{Re}\left(C_{\ket{0_\mathrm{s}}}(\beta_{\hat Y_0})\right)=e^{-\pi/2}.
            \end{align}
            In this example, the time-dependent offset is the same for opposite cardinal states $\ket{+\bar\mu}$ and $\ket{-\bar\mu}$, \textit{\textit{i.e.}} $f_{\ket{+\bar\mu}}(T)=f_{\ket{-\bar\mu}}(T)$. Therefore, taking the difference 
            \begin{align}
                \Delta\expval{\hat\mu_0}\equiv\expval{\hat\mu_0}_+-\expval{\hat\mu_0}_-,
            \end{align}
            removes the time-dependent offset $f_{\ket{\pm\bar\mu}}(T)$ without having to know its exact form. With QEC, given that there is no evidence of any time-dependent offset\,\cite{Sivak2023}, taking the difference is valid. Therefore, to make the analysis consistent between the different protocols, the difference of Pauli expectation values $\Delta\expval{\hat\mu_0}$ is considered both without and with QEC. 

        \subsubsection{Logical lifetime - Method A}
        \label{ssec:method_A}

            Method A is the same as in Ref\,\cite{Sivak2023}, up to the additional step of looking at the difference of Pauli expectation values. Under the assumption that opposite cardinal states have the same lifetime, we have, from Eq.\,\eqref{eq:time_dependence_Pauli},
            \begin{align}
                \Delta\expval{\hat\mu_0}(T)=\Delta\expval{\hat\mu_0}(0)e^{-T/T_{\ket{\bar\mu}}},
                \label{eq:Pauli_exp_value_decay_fit}
            \end{align}
            where $T_{\ket{\bar\mu}}$ is the lifetime of logical states $\ket{+\bar\mu}$ and $\ket{-\bar\mu}$. Equation\,\eqref{eq:Pauli_exp_value_decay_fit} is fitted to the data of $\Delta\expval{\hat\mu_0}(T)$ with $\Delta\expval{\hat\mu_0}(0)$ fixed, leaving the logical lifetimes $T_{\ket{\bar\mu}}$ as the only fitting parameters. In the case of Fig.\,3(c) of the main text, the value of $\Delta\expval{\hat\mu_0}(0)$ used in the fit corresponds to the value averaged across the three different protocols (without QEC and with QEC, default and optimized).

            The resulting fit of the logical fidelity, shown in Fig.\,2(d) and Fig.\,3(c) of the main text, is calculated with
            \begin{align}
                F_\mathrm{L}(T)=\frac{1}{2}+\frac{1}{12}\left(2\Delta\langle\hat X_0\rangle(T)+\Delta\langle\hat Y_0\rangle(T)\right),
            \end{align}
            where $\Delta\expval{\hat\mu_0}(T)$ are the fits of Eq.\,\eqref{eq:Pauli_exp_value_decay_fit} to the data of $\Delta\langle\hat X_0\rangle(T)$ and $\Delta\langle\hat Y_0\rangle(T)$. The logical lifetime $T_\mathrm{L}$ is calculated from the lifetimes $T_{|\bar X\rangle}$ and $T_{|\bar Y\rangle}$ with\,\cite{Sivak2023}
            \begin{align}
                T_\mathrm{L}=3\left(\frac{2}{T_{|\bar X\rangle}}+\frac{1}{T_{|\bar Y\rangle}}\right)^{-1},
            \end{align}
            which corresponds to having the logical decay rate $\gamma_\mathrm{L}=1/T_\mathrm{L}$ as the average of the decay rates of the cardinal states.

        \subsubsection{Logical lifetime - Method B}
        \label{ssec:method_B}

            In Method B, the logical lifetime is obtained by fitting the logical fidelity to an exponential decay with
            \begin{align}
               F_\mathrm{L}(T)=\frac{1}{2}+\left(F_\mathrm{L}(0)-\frac{1}{2}\right)e^{-T/T_\mathrm{L}},
            \end{align}
            where $F_\mathrm{L}(0)$ is the initial logical fidelity, which is fixed, leaving the logical lifetime as the only fitting parameter. As later discussed in Sec.\,\ref{ssec:comparison}, both methods lead to slightly different logical lifetimes, but consistent ratios when comparing protocols.

    \subsection{Quantum error correction}

        \subsubsection{Closed-loop optimizations}
        \label{ssec:closed-loop_QEC}

            For the quantum error correction protocol, two sets of parameters are optimized sequentially. First, the parameters related to the QEC protocol itself are optimized. As discussed in the main text, these parameters are the effective finite-energy parameter $\Delta_\mathrm{s\textbf{B}s}$, the scaling of the second small displacement, parameterized by the ratio $\left|\beta_{\mathrm{s}_2}\right|/\left|\beta_{\mathrm{s}_1}\right|$, and the state rotation per round $\Delta\theta_\mathrm{rd}$. The idle time $T_\mathrm{id}$ is fixed at $40\,\unit{\micro\second}$. Secondly, the ECD control parameters are optimized as described in Sec.\,\ref{ssec:closed-loop_ECD} with the parameters of the optimized QEC protocol.

            In both cases, the cost function is the logical infidelity $\mathcal{C}=1-F_\mathrm{L}$ after $4$ rounds, where $F_\mathrm{L}$ is evaluated from the numerical fit described in Sec.\,\ref{ssec:Pauli_measurement}. The optimizer used is \texttt{GaussianProcess} from Boulder Opal of Q-CTRL with $100$ iterations, plus an initial guess from a previous optimization.

\begin{figure*}[t]
                \centering
                \includegraphics[scale=0.81]{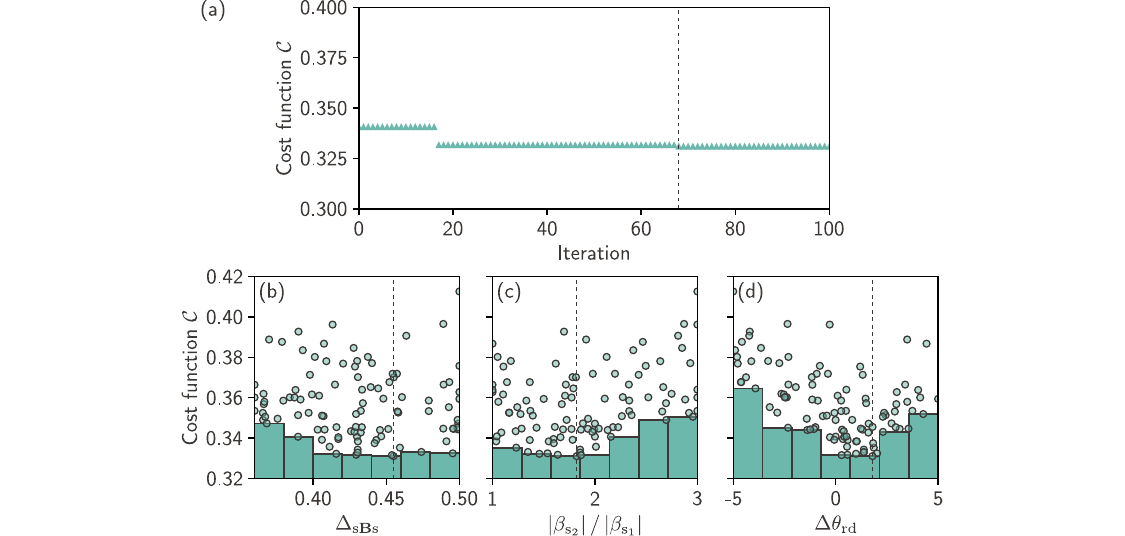}
                \caption{
                \textbf{Closed-loop optimization of QEC parameters.}
                (a)~Cumulative minimal value of the cost function $\mathcal{C}$ for the QEC protocol.
                Cost function $\mathcal{C}$ as a function of the (b) effective finite-energy parameter $\Delta_\mathrm{s\textbf{B}s}$, (c) scaling of the second small displacement $\left|\beta_{\mathrm{s}_2}\right|/\left|\beta_{\mathrm{s}_1}\right|$, and (d) state rotation per round $\Delta\theta_\mathrm{rd}$.
                Data points show the results from all iterations, while bars indicate the minimal value of the cost function within each bins.
                Dashed lines indicate the optimal value for each parameter.
                }
                \label{fig:closed-loop_optimization_QEC}
            \end{figure*}

            Results of the optimization for the parameters of the QEC protocol are shown in Fig.\,\ref{fig:closed-loop_optimization_QEC}(a). As seen in Fig.\,\ref{fig:closed-loop_optimization_QEC}(b), the optimization favors larger values of the effective finite-energy parameter, with a weak minima at $\Delta_\mathrm{s\textbf{B}s}=0.455$, with the default value at $\Delta_\mathrm{s\textbf{B}s}=\Delta=0.36$ for the default protocol. This observation indicates that, for the noise level of the current hardware, the finite-energy parameter maximizing the logical lifetime is probably higher than the value of $\Delta=0.36$ chosen for state preparation. The optimal ratio of $\left|\beta_{\mathrm{s}_2}\right|/\left|\beta_{\mathrm{s}_1}\right|=1.82$, with a default value at $1$ for the default protocol, is consistent with the value from Ref.\,\cite{Sivak2023}.

        \subsubsection{Measurement flow}

            Main results for the paper are those in Fig.\,3(c--d) of the main text for the following protocols: without QEC, with default QEC, and with optimized QEC. For these results, the measurement of Pauli expectation values are interleaved as follow:
            \begin{enumerate}
                \item For a given GKP state $\ket{\pm\bar\mu}$ with $\pm\bar\mu\in\left\{\pm\bar X,\pm\bar Y\right\}$ and the corresponding infinite-energy Pauli operator $\hat\mu_0\in\left\{\hat X_0,\hat Y_0\right\}$ and a given protocol, the Pauli expectation value $\expval{\hat\mu_0}_\pm=\expval{\pm\bar\mu|\hat\mu_0|\pm\bar\mu}$ is measured as a function of number of rounds $N_\mathrm{rd}$ from 0 to 9.
                \item Then the protocol is changed to cover without QEC, with default QEC, and with optimized QEC.
                \item The GKP state and corresponding Pauli operator is changed to cover $\langle\hat X_0\rangle_-$, $\langle\hat X_0\rangle_+$, $\langle\hat Y_0\rangle_-$, and $\langle\hat Y_0\rangle_+$.
            \end{enumerate}
            The whole procedure is repeated between $2$ and $4$ times. The error bars in Fig.\,3(c) of the main text are the $95\%$ confidence interval based on the standard deviation calculated from the different repetitions.

            Prior to any measurements of a Pauli expectation value, the auxiliary lifetime $T_{1\mathrm{q}}$ is measured with the storage mode at equilibrium. If the value is above a threshold value of $30\,\unit{\micro\second}$ (compared to an average value of $33\,\unit{\micro\second}$), the measurement of the Pauli expectation value is performed. If the value is below the threshold, the measurements of $T_{1\mathrm{q}}$ are repeated until three consecutive measurements are above the threshold value. Then, the Pauli expectation value measurement is performed.
            
            For every round with $N_\mathrm{rd}=0$, the readout visibility $\mathcal{V}$, reset error $\varepsilon_\mathrm{rt}$, and auxiliary equilibrium population $\bar n_\mathrm{q}^\mathrm{eq}$ are measured. As previously discussed, the average values from these measurements for $\bar n_\mathrm{q}^\mathrm{eq}$ and $\mathcal{V}$ are consistent with values used in simulation: $\bar n_\mathrm{q}^\mathrm{eq}=0.41\%$ instead of $0.38\%$ (Tab.\,\ref{tab:Lindbladian_parameters}) and $\mathcal{V}=0.9866$ instead of $0.9842$ (Sec.\,\ref{ssec:readout}). Furthermore, the reset error $\varepsilon_\mathrm{rt}$ is $1.3\%$ instead of $1.2\%$ for $N_\mathrm{ds}=2$.

        \subsubsection{Numerical simulations}

            The input states for the simulation of QEC protocols are the states obtained from numerical simulations of the initialization protocol described in Sec.\,\ref{sssec:init_simu}. Like for the initialization protocol, the numerical simulations are performed with Hilbert spaces sizes of $60$ and $3$ for the storage and auxiliary mode respectively. A size of $1$ is used for the resonator mode for free evolution, tracing out the mode from the initial state, and a size of $2$ is used for both QEC protocols, where auxiliary resets are simulated using this mode between QEC rounds.

            In the case of free evolution, the state is let to evolve freely under the undriven Lindblad master equation. The rotating phase of the storage mode is followed at its dressed frequency $\omega_\mathrm{s}$, considering that the auxiliary is mainly in its ground state following the reset at the end of state preparation.

            For the QEC protocols, the set of echoed conditional displacement is constructed following the method in Sec.\,\ref{sssec:ECD_simulations}, where the control parameters $\zeta$ and $c_\varphi$ are taken to optimize the ECD fidelity. The effective finite-energy parameter $\Delta_\mathrm{s\textbf{B}s}$ and the scaling of the second small displacement, parameterized by the ratio $\left|\beta_{\mathrm{s}_2}\right|/\left|\beta_{\mathrm{s}_1}\right|$, are set equal to the values used for the experimental protocols for both the default and the optimized QEC. The idle time between rounds is simulated just like free evolution.

            For each extracted time step, the logical information is extracted from a numerical simulation of the measurement of the real part of the characteristic function, $\mathrm{Re}(C(\beta_m))$, at a point $\beta_m$ corresponding to one of the infinite energy GKP logical Pauli operator. In accordance to the experimental methodology described in Sec.\,\ref{ssec:Pauli_measurement}, for each measured point $\beta_m$, we sweep its phase over a small interval in order to find the optimal phase corresponding to a local maximum (minimum) of $\mathrm{Re}(C(\beta_m))$ when measuring the $+1$ ($-1$) eigenstate of the operator. Similarly to the simulated characteristic function measurement for the initialization protocol (Sec.\,\ref{sssec:init_simu}), the auxiliary readout is not simulated and we directly compute $\expval{\hat\sigma_z}$, which we then correct for including readout errors.

        \subsubsection{Summary of the results}
        \label{ssec:comparison}

            \begin{table*}[t]
            \centering
            \begin{tabular}{l|l|l|l|ll|l}
            Encoding& Protocol & Method & Type & \multicolumn{2}{c}{Lifetime (ms)} & Relative difference (\%)\\
                    &          &        &      & Experiment      & Simulation      & \\
            \hline\hline
            GKP     & Free evolution & A & $\ket{\pm\bar X},\ket{\pm\bar Z}$ & $0.233(17)$ & $0.245(11)$ & $5.0(6)$ \\
                    &                &   & $\ket{\pm\bar Y}$                 & $0.127(11)$ & $0.138(08)$ & $8.0(1.2)$ \\
                    &                &   & Logical                           & $0.182(14)$ & $0.194(10)$ & $6.4(8)$ \\
                    &                & B & Logical                           & $0.196(11)$ & $0.208(08)$ & $5.7(6)$ \\
            \hline
                    & QEC, default   & A & $\ket{\pm\bar X},\ket{\pm\bar Z}$ & $0.224(08)$ & $0.262(13)$ & $14.5(1.2)$ \\
                    &                &   & $\ket{\pm\bar Y}$                 & $0.131(03)$ & $0.145(06)$ & $9.8(6)$ \\
                    &                &   & Logical                           & $0.181(06)$ & $0.206(09)$ & $12.3(9)$ \\
                    &                & B & Logical                           & $0.193(06)$ & $0.222(08)$ & $13.4(9)$ \\
            \hline
                    & QEC, optimized & A & $\ket{\pm\bar X},\ket{\pm\bar Z}$ & $0.266(21)$ & $0.323(23)$ & $17.8(2.7)$ \\
                    &                &   & $\ket{\pm\bar Y}$                 & $0.145(12)$ & $0.172(12)$ & $15.8(2.4)$ \\
                    &                &   & Logical                           & $\mathbf{0.208(17)}$ & $0.250(18)$ & $16.8(2.5)$ \\
                    &                & B & Logical                           & $0.224(19)$ & $0.271(21)$ & $17.2(2.8)$ \\
            \hline\hline
            Fock    & Free evolution & A & $\ket{\pm\bar X},\ket{\pm\bar Y}$ & ---         & $0.617(18)$ & ---         \\
                    &                &   & $\ket{\pm\bar Z}$                 & $0.336(07)$ & $0.336(07)$ & Simulation input \\
                    &                &   & Logical                           & ---         & $\mathbf{0.482(09)}$ & ---         \\
                    &                & B & Logical                           & ---         & $0.509(32)$ & ---         \\
            \end{tabular}
            \caption{
            \textbf{Summary of the state and logical lifetimes} obtained from experiment and simulation for the GKP and Fock qubits. The numbers in bold are the numbers explicitly given in the main text. 
            }
            \label{tab:lifetime_summary}
            \end{table*}

            Table\,\ref{tab:lifetime_summary} summarizes the logical lifetimes obtained in experiment and simulation for the three protocols and through methods A and B. The logical lifetime for the Fock encoding is obtained in simulation based on experimental parameters.
            
            Method B leads to systematically larger logical lifetimes. Notably, the logical lifetime for the Fock encoding in simulation obtained from method B of $0.509\,\unit{\milli\second}$ surpasses the upper bound of $\mathrm{Max}\left[T_\mathrm{L}^\mathrm{Fock}\right]=3T_{1\mathrm{s}}/2=0.504\,\unit{\milli\second}$ in the absence of any dephasing. This is despite the simulation having both intrinsic dephasing at rate $\kappa_{\phi\mathrm{s}}/2\pi=1.45$\,Hz and extrinsic dephasing from the auxiliary equilibrium population at approximate rate $\bar n_\mathrm{q}^\mathrm{eq}\kappa_{1\mathrm{q}}/2\pi=1.44$\,Hz. This would point out to method A being more valid, and this is why the values from that method are the ones mentioned explicitly in the main text.
            
            Nevertheless, as shown in Tab.\,\ref{tab:gain_summary}, the gain from QEC, from optimization, or from the Fock encoding, all calculated from ratios of different logical lifetimes, are very similar for both methods, indicating that such ratios are quite robust.

            \begin{table*}[t]
            \centering
            \begin{tabular}{l|l|l|ll}
            Parameter              & Protocol       & Method & Experiment   &   Simulation      \\
            \hline\hline
            Gain from QEC          & QEC, default   & A      & $1.00(11)$   & $1.06(10)$        \\
                                   &                & B      & $0.98(9)$    & $1.07(8)$        \\
            \hline
                                   & QEC, optimized & A      & $\mathbf{1.14(18)}$   & $1.29(15)$        \\
                                   &                & B      & $1.14(16)$   & $1.30(15)$        \\
            \hline\hline
            Gain from optimization & QEC            & A      & $\mathbf{1.15(13)}$   & $1.21(14)$        \\
                                   &                & B      & $1.16(13)$   & $1.22(14)$        \\
            \hline\hline
            Gain from Fock qubit   & Free evolution & A      & $0.377(30)$  & $0.403(21)$       \\
                                   &                & B      & $0.385(27)$  & $0.409(21)$       \\
            \hline
                                   & QEC, default   & A      & $0.375(12)$  & $0.428(20)$       \\
                                   &                & B      & $0.378(16)$  & $0.437(21)$       \\
            \hline
                                   & QEC, optimized & A      & $0.431(35)$  & $0.519(38)$       \\
                                   &                & B      & $0.440(43)$  & $0.532(48)$       \\
            \end{tabular}
            \caption{
            \textbf{Summary of the gains from QEC, from optimization, and from the Fock qubit.}
            The numbers in bold are the gain from QEC and optimization mentioned in the main text.
            }
            \label{tab:gain_summary}
            \end{table*}
    

%